\documentclass[12pt]{iopart} 
\usepackage{iopams}

\usepackage{color}
\usepackage{graphicx}
\usepackage{slashed}

\usepackage{pgfplots}

\newcommand{\ket}[1]{\left|#1\right\rangle}
\newcommand{\bra}[1]{\left\langle#1\right|}
\newcommand{\ii}{\mathrm{i}}
\newcommand{\ee}{\mathrm{e}}
\newcommand{\dd}{\mathrm{d}}
\newcommand{\eqref}[1]{(\ref{#1})\,}

\begin{document}
	\title{Relaxation dynamics of integrable field theories after a global quantum quench}
	\author{Emanuele Di Salvo and Dirk Schuricht}
	\address{Institute for Theoretical Physics, Center for Extreme Matter and Emergent Phenomena, Utrecht University, Princetonplein 5, 3584 CE Utrecht, The Netherlands}
	\ead{e.disalvo@uu.nl, d.schuricht@uu.nl}
	
	\begin{abstract}
	We apply the linked cluster expansion as well as the quench action approach to study the time evolution of one-point functions after a quantum quench in integrable field theories. We argue that the relaxation towards the stationary value fundamentally differs depending on the locality properties of the considered observable: while for local operators both exponential and power-law decaying terms are present in the dynamics, for semi-local operators the latter are absent. We explicitly confirm this for the Ising field theory, the sinh-Gordon model, and the repulsive sine-Gordon model. 
	\end{abstract}
	\vspace{2pc}
	
	\submitto{\JSTAT}
	\maketitle
	
	\section{Introduction}
	\label{Introduction}
Describing the many-body dynamics is a long-standing problem of theoretical physics since the early age of statistical mechanics, when works by Boltzmann and Gibbs elucidated most of the features enjoyed by ergodic classical systems. More recently, the dynamics of many-body quantum systems have been thoroughly analysed; while in principle a similar argument in terms of ergodic evolution can be carried on leading to the eigenstate thermalisation hypothesis ~\cite{Deutsch-91,Srednicki-99}, this case also shows exotic phenomena, as quantum scars~\cite{TurnerNature-18,TurnerPRB-18} and Hilbert space fragmentation~\cite{Sala-20}, that violate it (see,~\cite{Moudgalya-22} for a review). So far, ergodicity is the main mechanism that induces the system to relax towards a thermal state.

Another mechanism that prevents relaxation towards a thermal state is the presence of non-trivial conserved quantities. In integrable models, which possess an infinite number of these, the dynamics is highly constrained, such that the system is prohibited from exploring all energetically equivalent configurations, thus ergodicity is unable to drive the system to thermal equilibrium at late times~\cite{Rigol-07,Wouters-14,Pozsgay-14}. In addition, infinite number of conservation laws leads to stable excitations with completely elastic and fully factorisable scattering~\cite{Mussardo10}. This in turn hugely simplifies the analytic properties of the scattering matrix, opening the way for an analytic solution by the means of bootstrap procedures, including the form factors necessary for the computation of correlation functions~\cite{Smirnov92book}.

A further motivation for studying out-of-equilibrium integrable models comes from experimental realisation of such systems as cold atomic gases in optical traps. A direct way to generate out-of-equilibrium dynamics in these systems consists of a quantum quench~\cite{CalabreseCardy06}, ie, protocols where a system parameter is suddenly switched. In this setup a breakthrough came in the experiment of the ``quantum Newton cradle"~\cite{Kinoshita-06}, where the absence of relaxation in one-dimensional gases was observed. This sparked tremendous experimental activity, which lead to the observation of prethermalisation and the generalised Gibbs ensemble~\cite{Gring-12,Langen-15,Geiger-14,AduSmith-13,Kitagawa-11,Niklas-15,Meinert-15,Meinert-14,Malvania-21}.

Here we theoretically investigate the dynamics after a quantum quench in integrable systems. Specifically, using complementary analytic approaches we analyse the time evolution of one-point functions, which generically show exponentially decaying contributions as well as power-law tails. However, when the considered observable is semi-local with respect to the fundamental excitations of the model, power-law contributions are absent. 

In Section~\ref{Global quantum quenches in Integrable Field Theory}, we introduce the problem and we clarify the form and role of initial state after a global quench and the locality of operators in such theories. In Section~\ref{Main results}, we discuss our main result, which is formally derived in Section~\ref{sec:GA} by means of the linked cluster expansion and quench action methods. In Section~\ref{Discussion and conclusions}, we test our result for quenches in Ising, sinh-Gordon and repulsive sine-Gordon models. The technical calculations are presented in several appendices.

	\section{Global quantum quenches in integrable field theory}
	\label{Global quantum quenches in Integrable Field Theory}
\subsection{Integrable field theory}
In this work we consider a generic integrable field theory~\cite{Mussardo10}. It possesses an infinite number of conservation laws, thus allowing a full analytic treatment of the theory. Introducing the Faddeev--Zamolodchikov operators~\cite{ZamolodchikovZamolodchikov79,Faddeev80} $Z_a(\theta)$ and $Z^\dagger_a(\theta)$, which annihilate and create particles of type $a$ and rapidity $\theta$, the Hilbert space can be spanned by multi-particle states of the form
\begin{equation}
\ket{\theta_1,\ldots,\theta_N}_{a_1,\ldots,a_N}=Z_{a_1}^\dagger(\theta_1)\cdots Z_{a_N}^\dagger(\theta_N)\ket{0}
\end{equation}
where $\ket{0}$ denotes the vacuum. The momentum and energy are given in terms of the rapidities as
\begin{equation}
P=\sum_i m_{a_i}\sinh\theta_i,\quad E=\sum_i m_{a_i}\cosh\theta_i,
\end{equation}
with $m_{a_i}$ denoting the mass of a particle of type $a_i$, and we have set the velocity to unity. The infinite set of conservation laws ensures that scattering between different particles is factorised and elastic~\cite{Parke80}, which means that it can be fully described just in terms of the two-particle scattering matrix encoded in the algebra 
\begin{eqnarray}
&Z_a(\theta)Z_b(\theta') = S_{ab}^{cd}(\theta-\theta')Z_d(\theta')Z_c(\theta),\label{ZFA1}\\
&Z^{\dagger}_a(\theta)Z^{\dagger}_b(\theta') = S_{ab}^{cd}(\theta-\theta')Z_d^{\dagger}(\theta')Z_c^{\dagger}(\theta),\label{ZFA2}\\
&Z_a(\theta)Z_b^{\dagger}(\theta') = S_{bc}^{da}(\theta'-\theta)Z^{\dagger}_d(\theta')Z_c(\theta) + 2\pi\delta(\theta-\theta')\delta_{ab}.\label{ZFA3}
\end{eqnarray}
Moreover, the matrix elements of operators necessary to calculate correlation functions can be obtained using the form factor approach~\cite{Smirnov92book}. We give the list of axioms in~\ref{Form Factors}.

	\subsection{Local and semi-local operators}
	\label{Local and semi-local operators}
An important role in the time evolution of an operator expectation value after a quantum quench is played by the locality of such an operator with reference to the excitations of the model. When viewing a massive field theory as perturbation of a conformal field theory, local and semi-local operators can be specified by their operator product expansion with the fundamental field creating these excitations, which depends on the massive theory considered and the underlying conformal field theory~\cite{Ginsparg:notes, Cardy:08}. If the exponent of the diverging part is not an integer, the two operators have non-trivial braiding properties. This can be shown~\cite{YurovZamolodchikov91} to be related to the presence of a global symmetry; when this symmetry is broken, the different sectors are related exactly by the action of the semi-local operator. In this way, for example, different twisted boundary conditions can be implemented in conformal field theories. 

A consequence is that a semi-local operator is bound to connect different sectors of the broken symmetry, hence  
	\begin{equation}
		\label{Semi-localMEGeneral}
		_{a}\langle \mu|\tilde{\mathcal{O}}|\lambda\rangle_{a} = 0
	\end{equation}
for any pair of states $|\mu\rangle$ and $|\lambda\rangle$ that lie in the same sector $a$, and any semi-local operator\footnote{Sometimes we use the symbol $\tilde{\phantom{O}}$ to emphasise the semi-locality of the operator.} $\tilde{\mathcal{O}}$. In our case, a mechanism that breaks this symmetry is the finite-volume representation of the field theory. 
	
	As an example, we can consider the thermal deformation of the Ising model. The critical model is the minimal theory of massless Majorana fields $\psi,\bar{\psi}$ with central charge $c=1/2$. It contains three relevant operators: the identity $\mathbf{1}$, the energy operator $\epsilon\sim\bar{\psi}\psi$, and the order (disorder) operator $\sigma$ ($\mu$), whose behaviour depends on the studied phase (low- or high-temperature phase, respectively). Considering the massive theory in the ordered phase (for temperatures below the critical value), $\mu$ will create excitations (hence it is dubbed ``disorder operator"). The ``order operator"  $\sigma$ has the following operator product expansion with the disorder one~\cite{Kadanoff:71},
	\begin{equation}
		\label{OPEsigmamu}
		\sigma(z,\bar{z})\mu(0,0)\sim\frac{1}{\sqrt{2}|z|^{1/4}}\left(\ee^{\ii\pi/4}\sqrt{z}\psi(0) + \ee^{-\ii\pi/4}\sqrt{\bar{z}}\bar{\psi}(0)\right).
	\end{equation}
	From this one can extract the semi-locality factor $l(\sigma)=-1$, as winding of the order operator around the disorder one gives a factor of $-1$. By duality, one obtains the same result in the disordered phase by exchanging the two fields. The theory possesses a global $\mathbb{Z}_2$ symmetry; sending $\psi\to-\psi$ is equivalent to the action of braiding the two operators. The breaking of this symmetry allows to treat separatedly the Ramond (R) and the Neveu--Schwarz (NS) sectors in the evaluation of expectation values.
	
	The other two systems  that we will consider below are the sine-Gordon and sinh-Gordon models. They can both be viewed as perturbations of the free massless bosonic theory with central charge $c=1$, and they both possess $\mathbb{Z}_2$ symmetry by sending $\phi\to-\phi$. In the first case, the fundamental field is known via a fermionic basis construction~\cite{Smirnov92book}, which makes the derivation and discussion identical to the one for the thermal Ising field theory above. In particular, the vertex operator $\ee^{\ii \beta\phi/2}$ will be semi-local with locality index $l_\pm(\ee^{\ii \beta\phi/2}) = -1$. In contrast, for the sinh-Gordon model the fundamental field is not known. In~\ref{FFShGM} we will derive the form factors of the semi-local twist field $\tau$ and compute its scaling dimension $\Delta_\tau$ in order to identify it with an operator in the conformal theory.

\subsection{Global quantum quenches and initial state}
The quantum quench setup~\cite{CalabreseCardy06,Polkovnikov-11} we consider here consists of the preparation of the system in some initial state $\ket{\psi}$ and its subsequent time evolution under the Hamiltonian of some integrable field theory, where we analyse the dynamics of one-point functions $\langle \mathcal{O}(t)\rangle_\psi$ of local and semi-local operators. The analogy of this quench setup to a boundary problem in rotated Euclidean space time, as originally discussed by Calabrese and Cardy for conformal field theories~\cite{CalabreseCardy06}, motivates us to consider initial states of the form of squeezed coherent states~\cite{FiorettoMussardo10,Sotiriadis-12}, 
\begin{equation}
		\label{InitialState}
		|\psi\rangle = \exp\left\{\int_{0}^{\infty}\frac{\dd\theta}{2\pi}K^{ab}(\theta)Z_a^\dagger(-\theta)Z_b^\dagger(\theta)\right\}|0\rangle.
\end{equation}
We note that this form of initial states naturally appears in free theories~\cite{Cazalilla06,Rossini-10} after global changes of some parameter. Here the operators $Z_a^\dagger(-\theta)Z_b^\dagger(\theta)$ create zero-momentum pairs of excitations above the (post-quench) vacuum. In the corresponding boundary field theory states of this form are known to be compatible with the integrability~\cite{GhoshalZamolodchikov94} provided 
\begin{eqnarray}
& &K^{a_1c_1}(\theta_1)K^{c_2c_3}(\theta_2)
S^{a_2c_4}_{c_2c_1}(\theta_1+\theta_2)
S^{b_2b_1}_{c_3c_4}(\theta_1-\theta_2)\nonumber\\
&&\qquad =
K^{c_1b_1}(\theta_1)K^{c_2c_3}(\theta_2)
S^{b_2c_4}_{c_3c_1}(\theta_1+\theta_2)
S^{a_2a_1}_{c_2c_4}(\theta_1-\theta_2),\label{eq:BYBE}\\
&&K^{ab}(\theta)=S^{ab}_{cd}(2\theta)K^{dc}(-\theta).\label{eq:BUE}
\end{eqnarray}
We assume the validity of \eqref{eq:BYBE} and \eqref{eq:BUE} in the initial state \eqref{InitialState}. Furthermore, the amplitudes $K^{ab}(\theta)$ have to possess a suitable behaviour at large rapidities, which can for example be insured by including an extrapolation time~\cite{CalabreseCardy06,FiorettoMussardo10} $\tau_0$ via the factor $\ee^{-2m\tau_0\cosh\theta}$. Given that the Fadeev--Zamolodchikov operators create the eigenstates of the post-quench Hamiltonian, the time evolution starting from \eqref{InitialState} can in principle be obtained by expansion of the exponential. This approach is known as the linked cluster expansion and will be discussed in Section~\ref{Linked Cluster Expansion} below. We include more details on our choice of the initial state in~\ref{Initial states and pair structure}.
	
	\section{Main result}
	\label{Main results}
The time evolution of a generic one-point function of the operator $\mathcal{O}$ after a global quantum quench as described above will take the form 
	\begin{equation}
		\label{Main}
		\langle \mathcal{O}(t)\rangle_\psi = f(t)\,e^{-\Gamma t} + g(t) + \langle \mathcal{O}\rangle_\rho.
	\end{equation}
	Here $f(t)$ and $g(t)$ are regular functions on the real positive axis, $\Gamma$ a constant decay rate, and $\langle O\rangle_\rho$ the expectation value over the steady state $\ket{\rho}$ reached at infinite times. The functions $f(t)$ and $g(t)$ do not contain further exponential decay. We note that in principle various exponentially decaying contributions with different decay rates $\Gamma_i$ can appear, as is for example the case in the attractive regime of the sine-Gordon model~\cite{CS17, Horvath-18}. However, we will not consider such cases here.

We analyse \eqref{Main} for relativistic integrable field theories in the thermodynamic limit. We claim that in the case of a semi-local operator the function $g(t)$ identically vanishes, ie, the whole time evolution will show exponential relaxation. In contrast, for local operators both functions are non-vanishing\footnote{The exception is the energy operator $\epsilon$ in the Ising field theory, where $f(t)=0$.} and, in the late time regime, behave as
	\begin{equation}
		\label{LCase}
		 f(t)\sim t^{-\alpha_0}e^{2\ii m_0t},\qquad g(t)\sim t^{-\alpha_\rho}e^{2\ii m_\rho t}, \qquad t\to\infty.
	\end{equation}
Here $\alpha_0$, $m_0$, $\alpha_\rho$, and $m_\rho$ are related to the lowest number of excitations, over the post-quench ground state or over the steady state, respectively, with non-zero matrix elements of the operator $\mathcal{O}$. 

The absence or presence of polynomial tails at late times is the principal qualitative distinction between the out-of-equilibrium dynamics of local and semi-local operators. On the other hand, we also highlight the fact that exponential decay is present for local operators as well, although it can be neglected at late times. 

	\section{General arguments}\label{sec:GA}
	In this section we present general arguments supporting our claim based on the linked cluster expansion as well as quench action approach. In the next section we discuss the results for three specific models, namely the Ising field theory, sinh-Gordon model and sine-Gordon model. 
	
	\subsection{Linked cluster expansion}
	\label{Linked Cluster Expansion}
	We compute expectation values after a quantum quench with an initial boundary state (\ref{InitialState}). This is done by the means of the linked cluster expansion~\cite{Calabrese-12jsm1,Calabrese-12jsm2}, ie, expanding the initial state in powers of the function $K(\theta)$, which we consider small; this case is often referred to as small quench regime. It should be stressed that this approach, compared to perturbation theory~\cite{Delfino14,DelfinoViti17}, preserves non-adiabaticity of the quench, ie, the initial state is still extensive. In more physical terms, it means that we are still considering out-of-equilibrium processes and not linear responses at equilibrium, but the density of excitations over the post-quench ground state is small enough to allow the expansion.
	
	We set up the expansion as follows: Starting with the expectation value
	\begin{equation}
		\langle \mathcal{O}(t)\rangle_\psi =\frac{\bra{\psi}e^{\ii Ht}\mathcal{O}e^{-\ii Ht}\ket{\psi}}{\langle\psi\ket{\psi}},
	\end{equation}
	we formally expand both numerator and denominator in powers of $K$. Here and in the following we drop the indices that label the species of the excitations in order to keep the notation simple; for an example where they are restored, see \ref{Local Sine-Gordon operator's time evolution}. We also consider one particle mass only,
	\begin{eqnarray}
		\label{NumExp}
		\langle\psi| e^{\ii Ht} \mathcal{O} e^{-\ii Ht}|\psi\rangle&=& \sum_{M,N=0}^{\infty}\frac{1}{M!N!}\int_{0}^{\infty}\prod_{a=1}^{M}\frac{\dd\xi_a}{2\pi}K^*(\xi_a)\prod_{b=1}^{N}\frac{\dd\theta_b}{2\pi}K(\theta_b)\nonumber\\*
		& &\times \langle \xi_1,-\xi_1,\dots,\xi_M,-\xi_M|\mathcal{O}|-\theta_N,\theta_N,\dots,-\theta_1,\theta_1\rangle\nonumber\\*
		& &\times e^{2\ii mt\left(\sum_{a=1}^M\cosh{\xi_a}-\sum_{b=1}^N\cosh{\theta_b}\right)}\nonumber\\
		& =& \sum_{M,N=0}^{\infty} C_{M,N}^\mathcal{O}(t),
	\end{eqnarray}
	which defines the contributions $C_{M,N}^\mathcal{O}(t)$. The norm of the initial state is expanded as 
	\begin{eqnarray}
		\label{DenExp}
		\langle\psi|\psi\rangle &=&\sum_{N=0}^{\infty}\frac{1}{N!^2}\int_{0}^{\infty}\prod_{a=1}^{N}\frac{\dd\xi_a}{2\pi}\frac{\dd\theta_a}{2\pi}K^*(\xi_a)K(\theta_a)\nonumber\\*
		& &\qquad\times \langle \xi_1,-\xi_1,\dots,\xi_N,-\xi_N|-\theta_N,\theta_N,\dots,-\theta_1,\theta_1\rangle\nonumber\\
		&= &\sum_{N=0}^{\infty} Z_{2N}.
	\end{eqnarray}
	Thus we obtain the formal expansion
	\begin{equation}
		\label{InvDenExp}
		\frac{1}{\langle\psi|\psi\rangle} = 1-Z_2+Z_2^2-Z_4+O(K^6),
	\end{equation}
	which identifies the diverging parts in the infinite-volume limit, namely the linked clusters. Both numerator and denominator contain divergencies, but the expectation value does not; hence all divergencies have to cancel when the product between (\ref{NumExp}) and (\ref{InvDenExp}) is taken. We use a finite-size regularisation scheme to extract the divergent terms as being proportional to a positive power of the system size $L$ in the intermediate calculations. The initial state in finite volume $|\psi\rangle_L$ is derived in \ref{Finite volume theory} in terms of Bethe roots, see \eqref{FVInitialSt}. After taking the product, we obtain new expansion contributions $D^\mathcal{O}_{M,N}(t)$ which are finite as $L\to\infty$,
	\begin{equation}
		\label{LC2}
		\lim_{L\to\infty}\left[\frac{\langle\psi| \mathcal{O}(t) |\psi\rangle_L}{\langle\psi|\psi\rangle_L}\right] = \lim_{L\to\infty}\frac{\sum_{M,N=0}^{\infty} C^\mathcal{O}_{M,N}(t,L)}{\sum_{N=0}^{\infty} Z_{2N}(L)} = \sum_{M,N=0}^{\infty} D^\mathcal{O}_{M,N}(t),
	\end{equation}
	where we indicate the $L$-dependence of the finite-volume quantities. 
	
	The matrix elements in the numerator need further regularisation, for which we follow Smirnov~\cite{Smirnov92book}, see also Refs.~\cite{SE12,BSE14}, 
	\begin{equation}
		\label{SmirnovDec}
		\langle A|\mathcal{O}| B\rangle = \sum_{A=A_1\cup A_2}\sum_{B=B_1\cup B_2}d(B_2)S_{AA_1}S_{B_1B}\langle A_2|B_2\rangle\langle A_1+\ii\vec{\epsilon}\,|\mathcal{O}| B_1\rangle.
	\end{equation}
	Here $A$, $A_i$, and so on represent sets of rapidities and $S_{AA_1}$ and $S_{B_1B}$ denote the product of two-particle scattering matrices required to bring the Fadeev--Zamolodchikov operators creating the states in the correct order. The factor $d(B_2)$ keeps track of a possible semi-locality of the operator $\mathcal{O}$ with respect to the excitations in the set $B_2$. Furthermore, $\ii\vec{\epsilon}$ is a short-hand to denote that to all rapidities in the set $A_1$ a small imaginary part $\epsilon_i$ is added. 
	
	The term containing no elements in the set $A_2$ and $B_2$, and thus no $\delta$-functions, is called the fully connected contribution. Similarly, the term where the sets $A_1$ and $B_1$ are empty is called the fully disconnected piece. All other terms are considered to be semi-connected. Fully connected and semi-connected contributions contain poles in the form factor by virtue of the annihilation pole axiom. The contributions we obtain from regularising such singularities we dub the pole contributions.

	In the following we apply the linked cluster expansion to local and semi-local operators. The idea is to formally expand our prediction \eqref{Main} in powers of $K$, which is also contained in the rate $\Gamma$, and distinguish the terms by their time dependence~\cite{Calabrese-12jsm1,SE12}. In fact, expanding the exponential will give terms $\propto\Gamma t$ preceeded by combinatorial prefactors (as well as higher powers), which originate from the annihilation poles of the form factors as shown in \ref{Exponential decay from annihilation poles}. Thus in order to argue that the time evolution of local operators contains both exponential as well as power-law decay, we have to establish that (i) there exist terms $D_{M,N;\mathrm{conn}}(t)$ within $D_{M,N}(t)$ that get exponentiated, ie, the higher-order contribution $D_{M+2,N+2}(t)$ has the form $D_{M+2,N+2}(t)\propto-\Gamma t D_{M,N;\mathrm{conn}}(t)+\ldots$ (thus implying $f\neq 0$~\cite{Calabrese-12jsm1,SE12}), and (ii) that $D_{M,N}(t)$ also contains terms ${D}_{M,N;\mathrm{stat}}$ and ${D}_{M,N;\mathrm{reg}}(t)$ that do not appear with linear time dependence at higher order, ie, $D_{M+2,N+2}(t)$ does neither contain the term $-\Gamma t {D}_{M,N;\mathrm{stat}}$ nor $-\Gamma t {D}_{M,N;\mathrm{reg}}(t)$ (thus implying $g\neq 0$). Similarly, for semi-local operators we argue that for every term $D_{M,N;\mathrm{conn}}(t)$ within $D_{M,N}(t)$ that is not already growing in time, there exists a higher-order contribution $D_{M+2,N+2}(t)\propto-\Gamma t D_{M,N;\mathrm{conn}}(t)+\ldots$. Since in this case every term is exponentiated, it follows that $g=0$~\cite{Calabrese-12jsm1,SE12}. Below we focus on the diagonal contributions $D_{N,N}(t)$ to exhibit the behaviour described above.

	\subsubsection{Local operators}
	\label{Local operators}
	In this sub-section we consider local operators $\mathcal{O}$. They possess non-vanishing matrix elements only between states in the same sector and, for sake of simplicity, we can just consider the Ramond sector. The reason behind it is just that in the infinite volume limit, the expectation values over the two sectors  are identical (the results are independent from boundary conditions in such limit); hence
	\begin{eqnarray}
		\label{LOSplitting}
		\lim_{L\to\infty} \bra{\psi}\mathcal{O}(t)\ket{\psi}_L &= \frac{1}{2} \lim_{L\to\infty} \Bigl[\,_\mathrm{R}\bra{\psi}\mathcal{O}(t)\ket{\psi}_\mathrm{R} +_\mathrm{NS}\bra{\psi}\mathcal{O}(t)\ket{\psi}_\mathrm{NS}\Bigr] \\
		&  = \lim_{L\to\infty}{}_\mathrm{R}\bra{\psi}\mathcal{O}(t)\ket{\psi}_\mathrm{R} .
	\end{eqnarray}
	For the first terms in the linked cluster expansion we have the following regular expressions in finite volume $L$, 
	\begin{eqnarray}
		\label{RegCs}
		C^\mathcal{O}_{0,0}(t,L) &= &\langle 0|\mathcal{O}|0\rangle_L, \\
		C^\mathcal{O}_{0,2}(t,L) &= &\sum_{I\in \mathrm{R}}  K(I)\frac{F_2^\mathcal{O}(-\theta,\theta)}{\rho_1(\theta;L)} e^{-2\ii mt \cosh\theta}=\bigl[C^\mathcal{O}_{2,0}(t,L)\bigr]^*,\\
		C^\mathcal{O}_{2,2}(t,L) &= &\sum_{I\in \mathrm{R}} \sum_{J\in \mathrm{R}} K^*(I)K(J)\langle I,-I|\mathcal{O}|-J,J\rangle_L e^{2\ii mt(\cosh\xi - \cosh\theta)}.
	\end{eqnarray} 
	Here $F_N^\mathcal{O}(\theta_1,\ldots,\theta_N)=\bra{0}\mathcal{O}\ket{\theta_1,\ldots,\theta_N}$ denotes the $N$-particle form factor, see \ref{Form Factors}. The mapping between the Bethe numbers $I$ and $J$ and respective rapidities $\xi$ and $\theta$ is given by the Bethe equations, with $\rho_1(\theta;L)$ denoting the Jacobian of the relation $I\leftrightarrow\theta$; see~\ref{Finite volume theory} for the details. Considering states from same symmetry and parity sectors makes the presence of disconnected contributions possible in diagonal terms even when the system is at finite volume~\cite{PozsgayTakacs08-1,PozsgayTakacs08-2}. They contain terms proportional to the connected parts of form factors $F^\mathcal{O}_{N;\mathrm{conn}}(\theta_1,\dots,\theta_N)$, ie, regular parts of diagonal form factors extracted with the following procedure in the infinite volume~\cite{LeclairMussardo99},
	\begin{eqnarray}
		\label{ConnFormFactors}
		&&F_{N;\mathrm{conn}}^\mathcal{O}(\theta_1,\dots,\theta_N) \nonumber\\
		&&\quad = \mathcal{FP}\left\{\lim_{\{\epsilon_i\}\to 0^+}F_{2N}^\mathcal{O}(\theta_N+\ii\pi+\ii\epsilon_N,\dots,\theta_1+\ii\pi+\ii\epsilon_1,\theta_1,\dots,\theta_N)\right\},
	\end{eqnarray}
	in which the operation $\mathcal{FP}$ stands for discarding divergent contributions arising from annihilation poles and taking the finite part when regulators from (\ref{SmirnovDec}) are removed. We stress that the connected part $F^\mathcal{O}_{N;\mathrm{conn}}(\theta_1,\dots,\theta_N)$ depends on $N$ rapidities but is obtained from the $2N$-particle form factor. Using the finite-volume representation of the matrix elements and splitting diagonal terms from off-diagonal ones, we can write the last expression as 
	\begin{eqnarray}
		\label{C22FV}
		C^\mathcal{O}_{2,2}(t,L)&=& \langle 0|\mathcal{O}|0\rangle_L\sum_{I\in \mathrm{R}}|K(I)|^2  \nonumber \\
		&& \hspace{-2cm} 
		+ \sum_{I\in \mathrm{R}}|K(I)|^2 \frac{F^\mathcal{O}_{1;\mathrm{conn}}(\theta)}{\rho_1(\theta;L)} + \sum_{I\in \mathrm{R}}|K(I)|^2 \frac{F^\mathcal{O}_{1;\mathrm{conn}}(-\theta)}{\rho_1(\theta;L)} \nonumber\\
		&& \hspace{-2cm} 
		+\sum_{I\in \mathrm{R}}|K(I)|^2\frac{F_{2;\mathrm{conn}}^\mathcal{O}(-\theta,\theta)}{\rho_2(-\theta,\theta;L)} \nonumber\\
		&& \hspace{-2cm}
		+\sum_{I\neq J\in \mathrm{R}} \sum_{J\in \mathrm{R}} K^*(I)K(J) \frac{F^\mathcal{O}_4(\xi+\ii\pi,-\xi+\ii\pi,-\theta,\theta)}{\rho_1(\xi;L)\rho_1(\theta;L)}e^{2\ii mt(\cosh\xi - \cosh\theta)}.
	\end{eqnarray}
	The first contribution is due to the presence of the fully disconnected part of the matrix element and will be cancelled by contributions from the denominator (\ref{DenExp}). The second line contains semi-connected contributions and the third the fully connected stationary part of the form factor, which are both time-independent. We note that the latter is a contribution of order $O(1/L)$, since $\rho_2\propto L^2$, thus it can be neglected in the thermodynamic limit [while the former is of order $O(1)$]. Such loss of memory of the initial pair structure in the final state expectation value has already been described in~\cite{FiorettoMussardo10}. Finally, in the fourth line the infinite-volume form factor encodes all the information regarding off-diagonal terms (in the sense that $\xi\neq\theta$) and non-trivial time evolution. Such terms are always regular at finite volume; when the infinite-volume limit is taken, kinetic singularities as annihilation poles become important and must be treated separately. Now taking the large-volume limit and changing the sum over Bethe numbers into integrals over particles' rapidities (see \ref{Finite volume theory} for the exact Jacobian of the change of variables) we obtain
	\begin{eqnarray}
		\label{C22}
		C^\mathcal{O}_{2,2}(t,L\to\infty) &=& C_{0,0}^\mathcal{O}mL\int_{0}^{\infty}\frac{\dd\theta}{2\pi}|K(\theta)|^2\cosh\theta \nonumber\\
		&&\hspace{-30mm} + \int_{-\infty}^{\infty}\frac{\dd\theta}{2\pi}|K(\theta)|^2F^\mathcal{O}_\mathrm{1;conn}(\theta)  \nonumber\\
		&&\hspace{-30mm}+\int_{0}^{\infty}\frac{\dd\xi}{2\pi}\frac{\dd\theta}{2\pi}K^*(\xi)K(\theta) F_4^\mathcal{O}(\xi+\ii\pi,-\xi+\ii\pi,-\theta,\theta)e^{2\ii mt(\cosh\xi - \cosh\theta)}.
	\end{eqnarray}
	Here the first line contains the infinite-volume divergence, which will be cancelled by a contribution coming from the norm of the initial state (\ref{InvDenExp}). Furthermore, we note that since we are considering a local operator here, the four-particle form factor is regular at $\xi=\theta$ and the whole expression does not need further regularisation. We can thus find the second order contribution in the infinite volume,
	\begin{eqnarray}
		\label{D22local}
		D_{2,2}^\mathcal{O}(t) &=& \int_{0}^{\infty}\frac{\dd\xi}{2\pi}\frac{\dd\theta}{2\pi}K^*(\xi)K(\theta) F_4^\mathcal{O}(\xi+\ii\pi,-\xi+\ii\pi,-\theta,\theta)e^{2\ii mt(\cosh\xi - \cosh\theta)} \nonumber \\ 
		&& + \int_{-\infty}^{\infty}\frac{\dd\theta}{2\pi}|K(\theta)|^2F^\mathcal{O}_\mathrm{1;conn}(\theta)  \nonumber \\
		& = & D_\mathrm{2,2;conn}^\mathcal{O}(t) + D_\mathrm{2,2;stat}^\mathcal{O}.
	\end{eqnarray}
	Hence the result consists of a fully connected contribution as well as a stationary one and it agrees with the result previously found in literature for Euclidean time in a system with boundaries~\cite{KormosPozsgay10}. The latter is the one-particle contribution in the LeClair--Mussardo expansion~\cite{FiorettoMussardo10,LeclairMussardo99} of the stationary value $\langle\mathcal{O}\rangle_\rho$. In the case of semi-local operators the latter vanishes identically.
	
	The next term we consider in the expansion is $D^\mathcal{O}_{4,4}(t)$. In particular, we will show that it has the form 
	\begin{equation}
	D^\mathcal{O}_{4,4}(t)=-\Gamma t D_\mathrm{2,2;conn}^\mathcal{O}(t)+D^\mathcal{O}_{4,4;\mathrm{reg}}(t)+D_\mathrm{4,4;stat}^\mathcal{O}+\ldots,
	\label{eq:D44general}
	\end{equation}
	where the dots represent further contributions that structurally differ from the ones given explicitly. In the first term we will determine the rate $\Gamma=O(K^2)$, and thus show that the term $D_\mathrm{2,2;conn}^\mathcal{O}(t)$ obtained above gets exponentiated. Furthermore, by studying $D^\mathcal{O}_{6,6}(t)$ below we show that no term $\propto \Gamma t D^\mathcal{O}_{4,4;\mathrm{reg}}(t)$ appears at higher order, thus implying that $D^\mathcal{O}_{4,4;\mathrm{reg}}(t)$ contributes to the power-law decay $g$. We also determine the static contribution $D_\mathrm{4,4;stat}^\mathcal{O}$, which represents a higher-order contribution to the LeClair--Mussardo expansion. 
	
	When regularising the matrix element in the calculation of $C_{4,4}^\mathcal{O}(t,L)$ we obtain various disconnected and connected pieces. Specifically we will evaluate the terms
	\begin{equation}
	C_{4,4}^\mathcal{O}(t,L)=C^\mathcal{O}_{4,4;\mathrm{div}}(t,L)+C^\mathcal{O}_{4,4;\mathrm{stat}}(L)+C^\mathcal{O}_{4,4;\mathrm{pol}}(t,L)+C^\mathcal{O}_{4,4;\mathrm{reg}}(t,L),
	\end{equation}
 	which are the piece divergent in the infinite volume, a contribution to the stationary value, and the contribution from the annihilation poles. Furthermore, the regular term $C^\mathcal{O}_{4,4;\mathrm{reg}}(t,L)$ is time dependent and originates from form factors where one pair of ingoing rapidities equals one pair of outgoing ones. First, divergent contributions are obtained when the left and right states in the expectation value are identical (as before, ie, when $I_1=J_1$ and $I_2=J_2$), or when they have at least one Bethe number in common; with the result
	\begin{eqnarray}
		C^\mathcal{O}_{4,4;\mathrm{div}}(t,L) &=& C^\mathcal{O}_{0,0}(t,L) \left[\sum_{I\in \mathrm{R}}|K(I)|^4 - \left(\sum_{I\in \mathrm{R}}|K(I)|^2\right)^2\right] \nonumber\\
		&&+ C^\mathcal{O}_{2,2}(t,L)\sum_{I\in \mathrm{R}}|K(I)|^2.
	\end{eqnarray}
	This will be cancelled by the contributions coming from the denominator \eqref{InvDenExp}. In addition, there is a stationary contribution
	\begin{eqnarray}
		C^\mathcal{O}_{4,4;\mathrm{stat}}(L) &=& \frac{1}{4}\sum_{I_1\in \mathrm{R}}\sum_{I_2\in \mathrm{R}}|K(I_1)|^2|K(I_2)|^2\nonumber\\*
		&&\times\frac{\rho_2(-\theta_2,-\theta_1;L)}{\rho_4(-\theta_2,\theta_2,-\theta_1,\theta_1;L)}F^\mathcal{O}_\mathrm{2;conn}(\theta_1,\theta_2) + \dots,
	\end{eqnarray}
	where the dots stand for all the possible arrangements of the pair $(\pm\theta_1,\pm\theta_2)$ in the connected form factor. In the infinite-volume limit this becomes
	\begin{equation}
		\label{StTerms44}
		C^\mathcal{O}_{4,4;\mathrm{stat}}(L\to\infty) = \frac{1}{2}\int_{-\infty}^{\infty}\frac{\dd\theta_1}{2\pi}\frac{\dd\theta_2}{2\pi}|K(\theta_1)|^2|K(\theta_2)|^2 F^\mathcal{O}_\mathrm{2;conn}(\theta_1,\theta_2),
	\end{equation}
	which is the second-order contribution $D_\mathrm{4,4;stat}^\mathcal{O}$ in the LeClair--Mussardo series. The third term is dubbed 'regular' (its meaning will be clear in following discussion)
	\begin{eqnarray}
		C^\mathcal{O}_{4,4;\mathrm{reg}}(t,L) &=& \frac{1}{4}\sum_{I_1,I_2\in \mathrm{R}}\sum_{J_2\in \mathrm{R}}|K(I_1)|^2K^*(I_2)K(J_2)\nonumber\\*
		&&\hspace{-18mm}\times \frac{\rho_1(-\theta_1;L)\,F^\mathcal{O}_\mathrm{6;reg}(\theta_1+\ii\pi,\xi_2+\ii\pi,-\xi_2+\ii\pi,-\theta_2,\theta_2,\theta_1)}{\rho_3(-\theta_1,\theta_1,\theta_2;L)\rho_1(\xi_2;L)}+ \dots,
	\end{eqnarray}
	where the dots stand again for possible arrangements of the semi-disconnected configuration, in which two couples share the same Bethe number and the others do not. The label reg indicates the regular part of the form factor when the Laurent expansion is taken in the neighbourhood of an ingoing rapidity close to an outgoing one, as $\xi_1\to\theta_1$ in case of the previous configuration; see \ref{Exponential decay and power law behaviour from thermal Form Factors} for an explicit definition. In the infinite-volume limit this becomes
	\begin{eqnarray}
		\label{C44regpolt}
		&&C^\mathcal{O}_{4,4;\mathrm{reg}}(t,L\to\infty) = \int_{-\infty}^{\infty}\frac{\dd\theta_1}{2\pi}|K(\theta_1)|^2\int_{0}^{\infty}\frac{\dd\xi_2}{2\pi}\frac{\dd\theta_2}{2\pi}K^*(\xi_2)K(\theta_2) \nonumber\\*
		&&\qquad\times F^\mathcal{O}_\mathrm{6;reg}(\theta_1+\ii\pi,\xi_2+\ii\pi,-\xi_2+\ii\pi,-\theta_2,\theta_2,\theta_1)e^{2\ii mt(\cosh\xi_2-\cosh\theta_2)}\nonumber\\
		&&\equiv D^\mathcal{O}_{4,4;\mathrm{reg}}(t).
	\end{eqnarray}
	 By analysing below the next diagonal term in the expansion, $D_{6,6}^\mathrm{O}(t)$, we show that this term does not get exponentiated. Thus it generates a pure polynomial time dependence of the expectation value, ie, it contributes to $g(t)$ in the general expression \eqref{Main}. 
	
	Before doing so, we clarify how exponential decay is obtained by means of the linked cluster expansion. It is known~\cite{SE12,BSE14} that it originates from the annihilation poles of the form factor. There are two consequences: first, we can split singular contributions coming from poles of the form factors in their Laurent series (which we label as $F^\mathcal{O}_{N;\mathrm{sing}}$) from the regular ones  $F^\mathcal{O}_{N;\mathrm{reg}}$; second, in large but finite volume we can approximate the sum as an integral and get regular results by means of Sokhotski--Plemelij theorem~\cite{DiSalvo:23}. Using the regularisation (\ref{SmirnovDec}) with $\xi_i\to\xi_i+\ii\epsilon_i$ we obtain 
	\begin{eqnarray}
		\label{C44connected}
		\hspace{-20mm}C^\mathcal{O}_{4,4;\mathrm{pol}}(t,L) = \frac{1}{4}\sum_{I_1,I_2\in \mathrm{R}} \sum_{J_1,J_2\in \mathrm{R}} K^*(I_1)K^*(I_2)K(J_1)K(J_2)\nonumber\\*
		\hspace{-15mm}\times\frac{F_{8;\mathrm{sing}}^\mathcal{O}(\xi_1+\ii\pi+\ii\epsilon_1,-\xi_1+\ii\pi-\ii\epsilon_1,\xi_2+\ii\pi+\ii\epsilon_2,-\xi_2+\ii\pi-\ii\epsilon_2,-\theta_2,\theta_2,-\theta_1,\theta_1)}{\rho_4(-\xi_2,\xi_2,-\xi_1,\xi_1;L)\rho_4(-\theta_2,\theta_2,-\theta_1,\theta_1;L)}\nonumber\\*
		\hspace{-15mm}\times e^{2\ii mt(\cosh\xi_1 + \cosh\xi_2 - \cosh\theta_1 - \cosh\theta_2)}+\ldots,
	\end{eqnarray}
	where the dots contain the regular form factor $F_{8;\mathrm{reg}}^\mathcal{O}$. Taking the infinite-volume limit allows us to turn sums into integrals and consider the pole at $\xi_2=\theta_2$. The singular part of the form factor reads 
	\begin{equation}
		\hspace{-10mm}F_{8;\mathrm{sing}}^\mathcal{O}(\xi_1+\ii\pi+\ii\epsilon_1,\dots,\theta_1) = \frac{F_4^\mathcal{O}(\xi_1+\ii\pi+\ii\epsilon_1,-\xi_1+\ii\pi-\ii\epsilon_1,-\theta_1,\theta_1)}{(\xi_2-\theta_2+\ii\epsilon_2)^2},
	\end{equation}
	which implies, after the integral and the limits $\epsilon_i\to0$ are taken as prescribed by Sokhotski--Plemelij theorem,
	\begin{eqnarray}
		&C^\mathcal{O}_{4,4;\mathrm{pol}}(t,L\to\infty) = -\Gamma t\int_{0}^{\infty}\frac{\dd\xi_1}{2\pi}\frac{\dd\theta_1}{2\pi}K^*(\xi_1)K(\theta_1) \nonumber\\*
		&\qquad\times F_4^\mathcal{O}(\xi_1+\ii\pi,-\xi_1+\ii\pi,-\theta_1,\theta_1)e^{2\ii mt(\cosh\xi_1 - \cosh\theta_1)}.
	\end{eqnarray}
	We note that there are exactly $(2!)^2$ different ways to perform this computation and all of them must be summed up, cancelling the combinatorial prefactor. Picking the double pole involves taking the derivative of the exponential and makes the contribution linear in time; the constant showing up after this procedure reads
	\begin{equation}
		\label{Gamma}
		\Gamma = \frac{2m}{\pi}\int_{0}^{\infty}\dd\theta\,\big|K(\theta)\big|^2\,\sinh{\theta}.
	\end{equation}
	We can see that the factor multiplied by $-\Gamma t$ equals $D_\mathrm{2,2;conn}^\mathcal{O}(t)$ defined in \eqref{D22local}. In contrast, the stationary contribution $D_\mathrm{2,2;stat}^\mathcal{O}$ does not get resummed into an exponential decay, ie, it contributes to the expectation value over the representation state.	 Thus in total we arrive at \eqref{eq:D44general}.
	
	In order to see whether the term $D^\mathcal{O}_{4,4;\mathrm{reg}}(t)$ gets exponentiated, we have to check whether the part of $C_{6,6}^\mathcal{O}$ that grows linear in time contains a piece proportional to \eqref{C44regpolt}. To this end we can restrict ourselves to the contributions from the annihilation poles of the form factor. The procedure is the same as for $C_{4,4;\mathrm{pol}}^\mathcal{O}(t,L)$, except for the fact that now the required twelve-particle form factor $F_{12}^\mathcal{O}$ has two double poles. Picking both of them would retrieve a term proportional to $t^2 C_{2,2}^\mathcal{O}(t,L)$ we are not interested in. Hence we need to pick just one of the two double poles and keep the regular part of the other. The result in the infinite volume reads
	\begin{eqnarray}
		& C^\mathcal{O}_{6,6;\mathrm{pol},1}(t,L\to\infty) =-\frac{\Gamma t}{4}\int_{0}^{\infty}\frac{\dd\xi_1}{2\pi}\frac{\dd\xi_2}{2\pi}\frac{\dd\theta_1}{2\pi}\frac{\dd\theta_2}{2\pi}\nonumber\\*
		&\qquad \times K^*(\xi_1)K^*(\xi_2)K(\theta_1)K(\theta_2) e^{2\ii mt(\cosh\xi_1 + \cosh\xi_2 - \cosh\theta_1 - \cosh\theta_2)}\nonumber\\*
		&\qquad \times F_{8;\mathrm{reg}}^\mathcal{O}(\xi_1+\ii\pi,-\xi_1+\ii\pi,\xi_2+\ii\pi,-\xi_2+\ii\pi,-\theta_2,\theta_2,-\theta_1,\theta_1), 
	\end{eqnarray}
	where the additional subindex 1 indicates that we picked only one of the double poles. We see that this differs structurally from (\ref{C44regpolt}). Hence we conclude that $D^\mathcal{O}_{4,4;\mathrm{reg}}(t)$ does not get exponentiated (at least with the rate $\Gamma$), but rather contributes to the polynomial tail. We also see that the polynomial tail arises from regular parts of the form factors provided at least one pair of rapidities in both the ingoing and outgoing states agree. Namely, such contributions arise directly from regularisation of disconnected pieces of form factors in finite volume regime.
	
	We add that previous works applying the linked cluster expansion did not yield such polynomial contributions because they either focused on semi-local operators~\cite{SE12,BSE14} or considered the energy operator in the Ising field theory~\cite{FiorettoMussardo10}.  
	
	In our analysis we have focussed on the diagonal terms $D_{M,M}^\mathcal{O}(t)$. Considering off-diagonal terms does not add anything regarding the appearance of power-law contributions. However, off-diagonal terms give rise to additional oscillations, as indicated in \eqref{LCase}.
		
	\subsubsection{Semi-local operators}
	\label{Semi-local operators}
	For semi-local operators $\tilde{\mathcal{O}}$, regularisation is naturally provided by the finite volume and sector splitting (see \ref{Finite volume theory} for more details), given that matrix elements with ingoing and outgoing rapidities in the same sector vanish, cf. \eqref{Semi-localMEGeneral}. In particular, this implies for any semi-local operator 
	\begin{equation}
		\label{SemiLocalME}
		_\mathrm{R}\langle A|\tilde{\mathcal{O}}|B\rangle_\mathrm{R} =  _\mathrm{NS}\langle A|\tilde{\mathcal{O}}|B\rangle_\mathrm{NS} = 0.
	\end{equation}
	Note that we implicitly assume an underlying $\mathbb{Z}_2$-symmetry.
	When expectation values of semi-local operators are computed in finite volume, ground state splitting and their action on same sectors \eqref{SemiLocalME} restricts the analysis to two related contributions that equate in the infinite volume limit [cf (\ref{LOSplitting})]
		\begin{eqnarray}
			\label{SLOS}
			\lim_{L\to\infty}\langle\psi|\tilde{\mathcal{O}}|\psi\rangle_L &=& \frac{1}{2}\lim_{L\to\infty}\left[\,_\mathrm{NS}\langle\psi|\tilde{\mathcal{O}}|\psi\rangle_\mathrm{R} + \,_\mathrm{R}\langle\psi|\tilde{\mathcal{O}}|\psi\rangle_\mathrm{NS}\right] \nonumber\\
			&=&\lim_{L\to\infty}\,_\mathrm{NS}\langle\psi|\tilde{\mathcal{O}}|\psi\rangle_\mathrm{R}.
		\end{eqnarray}
	As we understood by studying local operators in the previous section, the absence of disconnected pieces in the finite-volume regime implies the absence of stationary and pure power-law contributions. This is the main result of the section and for completeness we directly show it in the specific setting.
	
	Semi-local operators have fewer fully regular contributions than local ones, by virtue of the annihilation pole axiom and sector splitting. For this reason, when finite volume is strictly imposed (i.e. we consider $1/L$ to be finite and not infinitesimally small), disconnected elements are absent. However, in the limit $1/L\to 0$ rapidities in different symmetry sectors become infinitesimally close, giving rise to contributions that eventually cancel denominator's ones. Since this requires the careful treatment of the different symmetry sectors, it is preferable to work with the infinite-volume regularisation (\ref{SmirnovDec}) directly. Specifically we find
	\begin{eqnarray}
		\label{RegCs'}
		&C^{\tilde{\mathcal{O}}}_{0,0}(t,L) = \,{}_\mathrm{NS}\langle 0|\tilde{\mathcal{O}}|0\rangle_\mathrm{R}, \\
		&C^{\tilde{\mathcal{O}}}_{0,2}(t,L) = \sum_{I\in \mathrm{R}}  K(I)\frac{F_2^{\tilde{\mathcal{O}}}(-\theta,\theta)}{\rho_1(\theta;L)} e^{-2\ii mt \cosh\theta}=\bigl[C^{\tilde{\mathcal{O}}}_{2,0}(t,L)\bigr]^*.
	\end{eqnarray}
	The first non-regular term is given by $C_{2,2}^{\tilde{\mathcal{O}}}$, whose form factor contains a double pole,
	\begin{eqnarray}
		\label{C22SL}
		&C^{\tilde{\mathcal{O}}}_{2,2} (t,L)=  \sum_{I\in \mathrm{R}} \sum_{J\in \mathrm{NS}} K^*(I)K(J)\bra{I,-I}\tilde{\mathcal{O}}\ket{-J,J}e^{2\ii mt(\cosh\xi - \cosh\theta)},\nonumber
	\end{eqnarray}
	and in the large but finite-volume regime we can extract the linear term in $L$ and pole contribution as usual~\cite{SE12,BSE14}. Here as well, extra care must be taken when (\ref{SmirnovDec}) is applied, given that also semi-disconnected terms (ie, terms that involve both disconnected and connected pieces) contain single pole contributions. They are cancelled by boundary contributions in Sokhotsky--Plemelij as well as in other cases~\cite{DiSalvo:23}. Its infinite-volume limit then reads
	\begin{eqnarray}
		\label{C22'}
		&C^{\tilde{\mathcal{O}}}_{2,2} (t,L)= mLC^{\tilde{\mathcal{O}}}_{0,0}(t,L)\int_{0}^{\infty}\frac{\dd\xi}{2\pi}|K(\xi)|^2 \cosh\xi - \Gamma t \,C^{\tilde{\mathcal{O}}}_{0,0}(t,L) \\
		& \quad+ \int_{0}^{\infty}\frac{\dd\xi}{2\pi}\frac{\dd\theta}{2\pi}K^*(\xi)K(\theta)F_{4;\mathrm{reg}}^{\tilde{\mathcal{O}}}(\xi+\ii\pi,-\xi+\ii\pi,-\theta,\theta)e^{2\ii mt(\cosh\xi - \cosh\theta)}. \nonumber
	\end{eqnarray}
	The first term on the right-hand side diverges in the large-volume limit but is cancelled by a corresponding contribution from the norm of the initial state. The term regular here has the same meaning as in (\ref{C44regpolt}). From the second line of (\ref{C22'}) we cannot extract any stationary term since the connected form factor of a semi-local operator is always vanishing. Thus we arrive at the second-order contribution to the linked cluster expansion of a semi-local operator, 
	\begin{equation}
		D^{\tilde{\mathcal{O}}}_{2,2} (t) = -\Gamma t \,C^{\tilde{\mathcal{O}}}_{0,0}(L\to\infty) + \dots \equiv -\Gamma t D^{\tilde{\mathcal{O}}}_{0,0} + D^{\tilde{\mathcal{O}}}_{2,2;\mathrm{conn}} (t),
	\end{equation}
	where the last term contains the form factor $F_{4;\mathrm{reg}}^{\mathcal{O}}$. We can also exclude that for higher order terms polynomial or stationary contributions are present: such terms arise from matrix elements containing equal rapidities in the ingoing and outgoing states, but they are absent for semi-local operators given (\ref{SemiLocalME}). 
	
	As an example, we consider the term $C_{4,4}^{\tilde{\mathcal{O}}}$, which reads in the finite-volume regularisation
	\begin{eqnarray}
		&C^{\tilde{\mathcal{O}}}_{4,4} (t,L)=  \frac{1}{4}\sum_{I_1,I_2\in \mathrm{R}} \sum_{J_1,J_2\in \mathrm{NS}} K^*(I_1)K^*(I_2)K(J_1)K(J_2)\\*
		&\times\bra{I_1,-I_1,I_2,-I_2}\tilde{\mathcal{O}}\ket{-J_2,J_2,-J_1,J_1}e^{2\ii mt(\cosh\xi_1 + \cosh\xi_2 - \cosh\theta_1 + \cosh\theta_2)}.\nonumber
	\end{eqnarray}
	Again, this contribution is finite and well-behaved in this scheme, as for (\ref{C22SL}).
	When $L$ is sent to infinity, disconnected contributions emerge and they must be split using (\ref{SmirnovDec}) and regularised via (\ref{DeltaReg}),
	\begin{equation}
		C^{\tilde{\mathcal{O}}}_{4,4;\mathrm{div}} (L)  = C^{\tilde{\mathcal{O}}}_{0,0}(t,L) Z_4- C^{\tilde{\mathcal{O}}}_{0,0}(t,L)Z_2^2+ C^{\tilde{\mathcal{O}}}_{2,2}(t,L)Z_2.
	\end{equation}
	These terms are correctly cancelled by the denominator (\ref{DenExp}) also in this limit. At this point we can study the analytic behaviour of the infinite-volume form factor in (\ref{C22SL}), which has double poles when $\xi_1\to\theta_1$ and $\xi_2\to\theta_2$ (and all other combinations thereof of two equal ingoing and outgoing rapidities). Three contributions are then present in the Laurent expansion: first, we pick the term containing two double poles; second, we pick the four terms containing one double pole; third, we consider the fully regular contribution. To the extent of this section, we just consider the second case, since the first leads to a quadratic correction to the linear term in time contained in (\ref{C22'}) and the third to a decaying contribution at late times. After taking the double pole contributions and having removed the semi-disconnected parts, we get
	\begin{eqnarray}
		C^{\tilde{\mathcal{O}}}_{4,4;\mathrm{pol},1}(t,L\to\infty) &=-\Gamma t\int_{0}^{\infty}\frac{\dd\xi}{2\pi}\frac{\dd\theta}{2\pi}K^*(\xi)K(\theta)e^{2\ii mt(\cosh\xi- \cosh\theta)}\nonumber\\*
		&\qquad\times F_{4;\mathrm{reg}}^{\tilde{\mathcal{O}}}(-\xi+\ii\pi,\xi+\ii\pi,-\theta,\theta). 
	\end{eqnarray}
	It can be seen that the power-law term obtained in (\ref{C22'}) has actually to be resummed and thus leads to exponentially damped contributions. At fourth order in the linked cluster expansion, we arrive to the following expression 
	\begin{eqnarray}
		D^{\tilde{\mathcal{O}}}_{4,4} (t) &=& - \Gamma t \int_{0}^{\infty}\frac{\dd\xi}{2\pi}\frac{\dd\theta}{2\pi}K^*(\xi)K(\theta)
		\nonumber\\*
		&&\qquad\times F_{4;\mathrm{reg}}^{\tilde{\mathcal{O}}}(-\xi+\ii\pi,\xi+\ii\pi,-\theta,\theta)e^{2\ii mt(\cosh\xi- \cosh\theta)} \nonumber\\
		&&+ \frac{\Gamma^2 t^2}{2}  \,C^{\tilde{\mathcal{O}}}_{0,0}(L\to\infty) + \dots \nonumber \\
		& \equiv& \frac{\Gamma^2 t^2}{2}D^{\tilde{\mathcal{O}}}_{0,0} -\Gamma t D^{\tilde{\mathcal{O}}}_{2,2;\mathrm{conn}}(t) + D^{\tilde{\mathcal{O}}}_{4,4;\mathrm{conn}}(t).
	\end{eqnarray}
	We can thus infer that the time evolution for semi-local operators is fully exponential, ie, we obtain $g(t)=0$ in the general expression \eqref{Main}.
	
	\subsection{Comparison with quench action method}
	\label{Comparison with Quench Action method's results}
	As we have seen in the previous part, computing the time evolution of one-point functions after a global quench requires to sum over the whole Hilbert space twice, which is an enormous computational effort (both for numerical and analytical methods). One way to reduce this complexity is the quench action method~\cite{CauxEssler13,Caux16}, which relies on the assumption that a steady state ('representative state') $|\rho\rangle$ exists whose properties can be determined by solving a generalised thermodynamic Bethe ansatz (gTBA) equation. As a consequence, the one-point function can be written as
	\begin{equation}
		\label{QAm}
		\langle \mathcal{O}(t)\rangle_\psi = \frac{\langle\psi| \mathcal{O}(t)|\rho\rangle}{2\langle\psi|\rho\rangle} + \mbox{h.c.},
	\end{equation}
	with the initial state $|\psi\rangle$. The above should be considered in a finite volume. By expanding the initial state in the eigenstates of the post-quench Hamiltonian $\{|\lambda\rangle\}$ one arrives at the following expression, 
	\begin{equation}
		\label{TimeEvol}
		\langle \mathcal{O}(t)\rangle_\psi = \frac{1}{2} \lim_{L\to\infty}\left[\sum_{|\lambda\rangle}e^{-\delta s_\lambda^* + \delta s_\rho + \ii(E_\lambda - E_\rho)t}\langle\lambda|\mathcal{O}|\rho\rangle_L + \mbox{h.c.}\right],
	\end{equation}
	which can be written in terms of Dirac $\delta$-functions and form factors of the operator $\mathcal{O}$ using (\ref{SmirnovDec}). In the above expression, $E_\lambda$ denotes the energy of the state $\ket{\lambda}$, while $\delta s_\lambda=-\ln\langle\lambda\ket{\psi}$ encodes the overlaps with the initial state. Picking the diagonal term with $\ket{\lambda}=\ket{\rho}$ immediately gives a stationary contribution. When analysing this expression, special care has to be given to the contributions from the poles of the appearing form factors. In the vicinity of these poles they can be split into a regular and a singular part in terms of multidimensional Laurent series. It is shown in \ref{Exponential decay from annihilation poles} how these poles together with the pair structure of the initial state $|\psi\rangle$ give rise to exponential decay with a universal inverse decay time given by
	\begin{eqnarray}
		\Gamma = \frac{2m}{\pi}\int_{0}^{\infty}\dd\theta\,\big|K(\theta)\big|^2\,\sinh{\theta} + O(K^3),
	\end{eqnarray}
	in agreement with \eqref{Gamma}. Thus the pole contributions give rise to the first part $f(t)\,e^{-\Gamma t}$ in our main result \eqref{Main}. The regular pieces of the form factors, on the other hand, provide the power-law contributions contained in $g(t)$, As we have shown in Section \ref{Linked Cluster Expansion} these are present when the operator has non-vanishing matrix elements in the same sector at finite volume (which are absent for semi-local operators). These can be described as well in terms of generalised hydrodynamics~\cite{Doyon-23} and thermal form factors (see \ref{Exponential decay and power law behaviour from thermal Form Factors}). If present, these power-law contributions describe the late-time dynamics of the system. In the language of generalised hydrodynamics such terms can be directly linked to the overlap of the operator with the conserved quantities, ie, $V^\mathcal{O}$~\cite{CuberoPanfil19,CuberoPanfil20}. The latter quantity is related to LeClair--Mussardo expansion of thermal expectation values~\cite{LeclairMussardo99}.
	
	These two kind of contributions in \eqref{Main} are distinguishable by the fact that the first, $f(t)\,e^{-\Gamma t}$, is determined by excitations over the post-quench ground state, while the second, $g(t)$, is given by excited states over the representative state (ie, the post-quench steady state). Given that the latter can be viewed as a 'sea of excitations above the ground state', the excitations above this steady state can be viewed as particle-hole excitations. This property is crucial when we consider matrix elements of semi-local operators, since they have vanishing overlap with the steady state and all its finite excitations; hence, one can distinguish local from semi-local operators just by investigating the late time dynamics: a power-law decay implies local operators while pure  exponential damping means semi-local ones. In~\ref{Examples: Ising and Sinh-Gordon models} we explicitly show for the Ising and sinh-Gordon models that the polynomial tails originate indeed from excitations over the representative state.

	\section{Discussion of specific models}
	\label{Discussion and conclusions}
	The goal of this section is to compare our general claim to results obtained for specific integrable field theories. We both consider free and interacting systems.
	
	\subsection{Thermal Ising field theory}
	\label{Thermal Ising field theory}
	The thermal Ising field theory arises from Ising conformal field theory when deformed by its energy operator $\varepsilon$. At the conformal point, the model describes the dynamics of two decoupled Majorana fields. The deformation couples the two Majorana fields via the mass term, describing the scaling limit of the Ising model close to the critical point at zero magnetic field. The model is still a free system also in the massive regime. The action is determined by a Dirac Lagrangian for fields generating massive excitations
	\begin{equation}
		\label{IsingL}
		\mathcal{L} = \Psi^\dagger(\ii\slashed{\partial} - m)\Psi.
	\end{equation}
	Stable particles are kinks interpolating between two different vacua of the corresponding Landau-Ginzburg effective theory with fermionic exchange rule, ie, $S=-1$. We consider two different relevant operators, namely $\varepsilon$ and $\sigma$; the first one is local while the second is semi-local, as already mentioned in Section~\ref{Local and semi-local operators}. In the related lattice model, these two represent respectively the exchange energy between two neighbouring spins and the total magnetisation of the chain~\cite{Calabrese-12jsm1,Calabrese-12jsm2}. We are implicitly assume the system to be in the ordered phase, ie, we consider the disorder operator $\mu$ to be creating stable excitations above the vacuum. However, this construction can be mapped to the disordered phase by Kramers--Wannier duality, where excitations are now created by the order operator and the disorder operator plays the role of measuring the magnetisation of the system.
	
	The quench we are considering consists of suddenly changing the mass of the particles $m_0\to m$. The initial state is then exactly of the form of the boundary squeezed state (\ref{InitialState}) with the amplitude $K(\theta)$ being explicitly known~\cite{Sotiriadis-12,Rossini-10}. The simplicity of the form factors of the energy operator then allows a straightforward computation of the time evolution by the means of linked cluster expansion~\cite{FiorettoMussardo10,Calabrese-12jsm1} and quench action method (see \ref{Examples: Ising and Sinh-Gordon models}) with the result
	\begin{equation}
		\label{Epsilon}
		\langle\varepsilon(t)\rangle_\psi  = \langle\varepsilon(0)\rangle_\rho + \mathfrak{Re}\left\{\int_{0}^{\infty}\frac{\dd\theta}{2\pi}\frac{m\sinh\theta K(\theta)}{1+|K(\theta)|^2}e^{2\ii mt\cosh\theta}\right\}.
	\end{equation}
	The result for the magnetisation operator can be retrieved as well, it reads~\cite{Calabrese-12jsm1,SE12, CauxEssler13,Granet-20}
	\begin{equation}
		\label{Sigma}
		\langle\sigma(t)\rangle_\psi  = \bar{\sigma} \left[1+\frac{\Omega}{mt} - \frac{1-\frac{m}{m_0}}{8\sqrt{\pi}}\frac{\cos(2mt-\frac{\pi}{4})}{mt^{3/2}} + \dots\right] e^{-\Gamma t} + \dots,
	\end{equation}
	where $\bar{\sigma}=\bra{0}\sigma\ket{0}$ denotes the expectation value in the post-quench vacuum state, the dots represent corrections to the prefactor as well as terms decaying faster than $e^{-\Gamma t}$, and 
	$\Omega$ is an uninteresting constant.
	
	These two expectation values already contain most of the features described beforehand: the local operator $\varepsilon$ has non-vanishing overlap (and thus expectation value) with the steady state and shows power-law decay. That additional exponentially decaying terms are absent is a peculiar feature of free theories. On the contrary, the semi-local operator $\sigma$ does not relax towards a non-zero one-point expectation value and shows exclusive exponential decay, governed by the relaxation rate $\Gamma$. Both expectation values show oscillations with frequency $\sim 2m$, which is the energy required to excite a pair of excitations above the post-quench vacuum. Thus these results are consistent with (\ref{Main}).
	Moreover, the main claim (\ref{Main}) agrees with expectations coming from thermal two-point functions in XY spin chain in presence of perpendicular magnetic field too~\cite{Its93}.
	
	\subsection{Sinh-Gordon field theory}
	\label{Sinh-Gordon field theory}
	Arguably the simplest interacting integrable relativistic field theory is the sinh-Gordon model: it contains one bosonic field $\phi$ that interacts with itself via a hyperbolic cosine potential; the Lagrangian reads
	\begin{equation}
		\mathcal{L} = \frac{1}{2}(\partial\phi)^2 - \frac{\mu^2}{g^2}(\cosh{g\phi} -1).
		\label{eq:ShGM}
	\end{equation}
	The parameters $\mu$ and $g$ are respectively the bare mass and the coupling constant, which are linked to the physical mass $m$ and the effective coupling $B$ as
	\begin{equation}
		m^2 = \mu^2\frac{\sin\pi B}{\pi B}, \quad B = \frac{g^2}{8\pi + g^2}.
	\end{equation}
	The stable excitations are spinless, neutral particles of mass $m$ with the two-particle scattering matrix 
	\begin{equation}
		S(\theta) = \frac{\sinh\theta - \ii\sin\pi B}{\sinh\theta + \ii\sin\pi B},
		\label{eq:ShGMSmatrix}
	\end{equation}
	which is invariant under the duality transformation $g\to\sqrt{8\pi}/g$, a symmetry that is not present at the level of the Lagrangian \eqref{eq:ShGM}, see Ref.~\cite{Konik-21} for a more detailed discussion. For clarity, we will avoid the self-dual point $g=\sqrt{8\pi}$ and restrict ourselves to the regime $0<g<\sqrt{8\pi}$.
	
	One of the most natural operators are the so-called vertex operators $e^{\alpha g\phi}$, which contain information regarding any power of the field operator. The vertex operators are labelled by the real number $\alpha$. They are local and their time evolution was (partially) computed in~\cite{DiSalvo:23} in the small quench regime,
	\begin{eqnarray}
		\label{VO}
		\langle e^{\alpha g\phi(t)}\rangle_\psi  &=&\mathcal{G}_\alpha + \int_0^{\infty}\frac{\dd\theta}{2\pi}|K(\theta)|^2F_\mathrm{1;conn}^\alpha(\theta)\nonumber\\
		&& + 2e^{-\Gamma t}\,
		\mathfrak{Re}\int_0^{\infty}\frac{\dd\theta}{2\pi}K(\theta)F_2^\alpha(-\theta,\theta)e^{-2\ii mt\cosh{\theta}} \nonumber\\
		&& + \mathfrak{Re}\int_{0}^{\infty}\frac{\dd\theta}{2\pi}\int_{0}^{\infty}\frac{\dd\xi}{2\pi}K(\theta)|K(\xi)|^2\nonumber\\*
		&&\qquad\qquad\times F_{4;\mathrm{reg}}^\alpha(\xi+\ii\pi,-\theta,\theta,\xi)e^{-2\ii mt\cosh\theta},
	\end{eqnarray}
	where $\mathcal{G}_\alpha=\bra{0}e^{\alpha g\phi}\ket{0}$~\cite{LukyanovZamolodchikov97}. Please note that in the previous work~\cite{DiSalvo:23} the regular piece was not considered. As compared to the energy operator in the Ising field theory, we observe exponential decay, power-law decay, and oscillations, ie, all contributions expected from \eqref{Main}. The full derivation, which includes the last line, can be carried out both using the linked cluster expansion or thermal form factors; it is contained in \ref{Examples: Ising and Sinh-Gordon models}.

	Defining a semi-local operator for the sinh-Gordon model requires finding a semi-local one in the related conformal field theory, which is given by the free boson or Gaussian model. In analogy with Ising model, we consider the twist operator $\tau$. In \ref{FFShGM} we derive its form factors in the sinh-Gordon model. As a check, we compute scaling dimension of the field, which agrees with the prediction from conformal field theory. Moreover, computing its time evolution (details in \ref{Semi-local Sinh-Gordon operator's time evolution}), we get
	\begin{equation}
		\label{Tau}
		\langle\tau(t)\rangle_\psi  = (\bar{\tau} + \dots)e^{-\Gamma t} ,
	\end{equation}
	where $\bar{\tau}$ again denotes the vacuum expectation value. The dots represent rapidly oscillating terms and higher powers in the function $K$ for small quenches. Thus again we find agreement with our hypothesis (\ref{Main}).
	
	\subsection{Sine-Gordon field theory}
	\label{Sine-Gordon field theory}
	A prototypical example of an integrable field theory is the famous sine-Gordon model; it emerges in many different contexts, such as low-energy spin chain realisations~\cite{Essler:Review}, trapped ultra-cold atoms~\cite{Giamarchi04} and, in its boundary version, chiral channel tunnelling~\cite{Fendley:95}. Its fermionised counterpart is the zero-charge sector of Thirring model~\cite{KorepinBogoliubovIzergin93} as first realised by Coleman~\cite{Coleman:75}. The model is defined by the Lagrangian
	\begin{equation}
		\mathcal{L} = \frac{1}{2}(\partial\phi)^2 - \frac{\mu^2}{\beta^2}(\cos{\beta\phi} -1).
	\end{equation}
	Even though the sine-Gordon and sinh-Gordon models are formally related by an analytic continuation in the interaction parameter, $g\to\ii\beta$, their physical properties are rather different. The sine-Gordon model possesses stable topological excitations, named solitons and antisolitons, which classically interpolate between the minima of the cosine potential. In addition, in the attractive regime, $\beta^2<4\pi$, there exist stable bound states. In the quantum theory this is highlighted by the presence of poles in the scattering matrix between solitons and antisolitons. However, here we consider the repulsive regime $4\pi<\beta^2<8\pi$ only. 
	
	Depending on the parameter $\alpha$, the vertex operators $e^{\ii\alpha\phi}$ possess different locality. Specifically, $e^{\ii\beta\phi}$ is local with respect to the fundamental field creating solitons or antisolitons, while $e^{\ii\beta\phi/2}$ has semi-locality index $l_a(e^{\ii\beta\phi/2})=-1$, where $a=\pm$. The time evolution of the latter was studied~\cite{BSE14} in the repulsive regime with the result
	\begin{equation}
		\label{SGbetahalf}
		\langle e^{\ii\beta\phi(t)/2}\rangle_\psi = \mathcal{G}_{\beta/2} e^{-\Gamma t} + \dots,
	\end{equation}
where the dots are containing higher order terms in the small quench approximation. In addition, in \ref{Local Sine-Gordon operator's time evolution} we determine the time evolution of the local vertex operator to be
	\begin{eqnarray}
		\label{SGbeta}
		\langle e^{\ii\beta\phi(t)}\rangle_\psi &=& \mathcal{G}_\beta + \int_0^{\infty}\frac{\dd\theta}{2\pi}K^{ab}(\theta)^*K_{ab}(\theta)F_{1,c\mathrm{;conn}}^\beta(\theta)\nonumber\\
		&&+ \int_{0}^{\infty}\frac{\dd\theta}{2\pi}K^{ab}(\theta)F_{2,ab}^\beta(-\theta,\theta)\nonumber \\
		&& + e^{-\Gamma t}\int_{-\infty}^{\infty}\frac{\dd\xi}{2\pi}\frac{\dd\theta}{2\pi}K^{ab}(\xi)^*\,K^{cd}(\theta)\nonumber\\*
		&&\qquad\times F^\beta_{4,\bar{a}\bar{b}cd,\mathrm{reg}}(\xi+\ii\pi,-\xi+\ii\pi,-\theta,\theta)e^{2\ii mt(\cosh\xi - \cosh\theta)}  \nonumber\\
		&& +  \int_{0}^{\infty}\frac{\dd\xi}{2\pi}\frac{\dd\theta}{2\pi}K^{ab}(\xi)^*K_{ab}(\xi)K^{cd}(\theta)\nonumber\\*
		&&\qquad\times  F^\beta_{4,\bar{a}acd;\mathrm{reg}}(\xi+\ii\pi,-\theta,\theta,\xi)e^{2\ii mt\cosh\theta} + \dots,
	\end{eqnarray}
	where the dots stand for higher order contributions in powers of $K$, and summation over the indices is implicit (and $\bar{a}=-a$). This result, as its counterpart in the sinh-Gordon theory, contains both exponential and power-law terms; both $f(t)$ and $g(t)$ scale as $t^{-3/2}$ in at late times. Thus our findings are in line again with (\ref{Main}).
	
	As is well-known, the sine-Gordon model effectively describes the low-energy properties of many lattice models. Thus the question arises whether this also applies to the quench setup. Numerical simulations~\cite{Barmettler-09,Barmettler-10} of the dynamics of the staggered magnetisation in the gapped phase of the XXZ Heisenberg chain give an affirmative answer in the sense that the staggered magnetisation is linked via bosonisation to the semi-local operator $\cos(\beta\phi/2)$ and the numerical data show pure exponential decay. We stress, however, that various limitations have to be considered: (i) lattice simulations are necessarily limited to finite times, thus one cannot rule out that power-law decay will eventually show up, (ii) the link between the initial lattice state and the squeezed state \eqref{InitialState} is unclear, (iii) the finite energy density introduced by the quench pushes the system away from the necessary low-energy regime, (iv) thus irrelevant perturbations dropped in the sine-Gordon model may affect the results, and (v) the definition of semi-locality of an operator may differ between lattice and field theory due to the differing reference states.
	
	Finally, let us briefly comment on more general semi-local operators $e^{\ii\alpha\phi}$ with $\alpha\neq\beta/2$. We first recall that a semi-classical treatment~\cite{KormosZarand15} yields a non-vanishing stationary expectation value $\langle e^{\ii\alpha\phi}\rangle_\rho$. In contrast, the quench action approach predicts a vanishing one for the reason that $e^{\ii\alpha\phi}$ connects different sectors in the finite volume theory; implying the vanishing of expectation values within the sectors, cf.  (\ref{Semi-localMEGeneral}). We attribute the discrepancy to the purely reflective scattering in the semi-classical approximation, which seems to prohibit relevant decay channels, as explicitly shown in~\cite{MocaKormosZarand15}. We note in addition that in the attractive regime ($\beta^2<4\pi$) the semi-classical approach seems to miss oscillations originating from the presence of zero-momentum breathers~\cite{CS17,Horvath-18}.
		
	\section{Conclusions}
	\label{Conclusions}
	In this paper we argued that the time evolution of one-point functions in integrable field theories after a global quantum quench is given by (\ref{Main}). Importantly, the distinction between local and semi-local operators concerns the presence of power-law terms; ie, while for local operators one generically has both exponential and power-law decay ($f\neq 0\neq g$), for semi-local ones the latter is absent ($g=0$). We have obtained these results using a combination of approaches to study the quench dynamics in integrable field theories, ie, the linked cluster expansion, quench action method, and thermal form factors. Moreover, these results extend to non-relativistic integrable field theories like the Lieb--Liniger model in repulsive and attractive regimes~\cite{DiSalvo:23}.
	
	Further steps could be the analysis of field theories containing bound states~\cite{CS17,Horvath-18} and to integrable lattice models. Our expectations are that our results extend to these type of models as well, although a refinement of our methods may be necessary.  A promising role might be played by generalised hydrodynamics~\cite{Bertini:16, CastroAlvaredo:16}; even though we did not apply this method here, generically hydrodynamic predictions allow us to infer power-law behaviour (as in \ref{Examples: Ising and Sinh-Gordon models}), and recently~\cite{Doyon:23} a way of characterising the exponential decay for local operators in this framework was established. 
		
	\section*{Acknowledgements}
	We thank Olalla Castro-Alvaredo, Jean-S\'{e}bastien Caux,  Axel Cort\'{e}s-Cubero, Benjamin Doyon, Fabian Essler, Jacopo De Nardis, Miłosz Panfil, Stefano Scopa and Andrew Urichuk for useful discussions. 
	
	\appendix
	\section{Form factor axioms}
	\label{Form Factors}
Using the Faddeev--Zamolodchikov algebra \eqref{ZFA1}--\eqref{ZFA3} one can exchange particles in the  asymptotic states, thus we can use the convention to order rapidities in both the in- and out-states in an increasing fashion. Now we can define form factors of an operator $\mathcal{O}$ as the matrix elements
		\begin{eqnarray}
			\label{FFDef}
			F^\mathcal{O}_N(\theta_1,\dots,\theta_N)_{a_1,\dots,a_N} 
			&=& \langle 0 |\mathcal{O}|\theta_1,\dots,\theta_N\rangle_{a_1,\dots,a_N}\\
			&=& \langle 0 |\mathcal{O}Z_{a_1}^\dagger(\theta_1)\dots Z_{a_N}^\dagger(\theta_N)\ket{0}
		\end{eqnarray}
		with $\theta_1<\theta_2<\ldots<\theta_N$. When just one particle species is present, we drop the indices of the form factors and scattering matrices. The form factors satisfy the following relations~\cite{Smirnov92book,Delfino04}:		
		\begin{itemize}
		\item Analyticity: The form factors are meromorphic functions in the physical strip $0\le\mathfrak{Im}\theta_N\le 2\pi$.
		\item Scattering axiom:
		\begin{eqnarray}
			\label{ScattAx}
			&&F^{\mathcal{O}}_{N}(\theta_1,\dots,\theta_i,\theta_{i+1},\dots,\theta_N)_{a_1,\dots,a_i,a_{i+1},\dots,a_N} \nonumber\\
			&&= S(\theta_i - \theta_{i+1})_{a_i,a_{i+1}}^{b_{i},b_{i+1}}F^{\mathcal{O}}_{N}(\theta_1,\dots,\theta_{i+1},\theta_i,\dots,\theta_N)_{a_1,\dots,b_{i+1},b_i,\dots,a_N}.
		\end{eqnarray}
		\item Periodicity axiom:
		\begin{eqnarray}
			\label{PeriodAx}
			&&F^{\mathcal{O}}_{N}(\theta_1+2\pi\ii,\theta_2,\dots,\theta_N)_{a_1,\dots,a_N} \nonumber\\
			& &= l_{a_1}(\mathcal{O})F^{\mathcal{O}}_{N}(\theta_2,\dots,\theta_N,\theta_1)_{a_2,\dots,a_{N},a_1} ,
		\end{eqnarray}
		where $l_a(\mathcal{O})$ represents the mutual semi-locality factor between the operator $\mathcal{O}$ and the fundamental field creating an excitation of type $a$.
		\item Lorentz transformation:
		\begin{equation}
			\label{LorentzTr}
			F^{\mathcal{O}}_{N}(\theta_1+\Lambda,\dots,\theta_N+\Lambda)_{a_1,\dots,a_N} =\ee^{s(\mathcal{O})\Lambda}F^{\mathcal{O}}_{N}(\theta_1,\dots,\theta_N)_{a_1,\dots,a_N} ,
		\end{equation}
		where $s(\mathcal{O})$ is the Lorentz spin of the operator. All operators considered here are spinless, thus $\ee^{s(\mathcal{O})}=1$.
		\item Annihilation pole axiom:
		\begin{eqnarray}
			\label{AnnPol}
			&&\mbox{Res}\left[F^{\mathcal{O}}_{N+2}(\theta',\theta,\theta_1,\dots,\theta_N)_{a,b,a_1,\dots,a_N} ,\theta'=\theta+\ii\pi\right] \nonumber\\ 
			&&=\ii C_{a,c}F_N^{\mathcal{O}}(\theta_1,\dots,\theta_N)_{b_1,\dots,b_N}
			\Bigl(\delta_{b_1}^{a_1}\dots\delta_{b_N}^{a_N}\delta_{b}^{c}\nonumber\\*
			&&\quad -l_a(\mathcal{O})S^{c_1,b_1}_{b,a_1}(\theta-\theta_1)S^{c_2,b_2}_{c_1,a_2}(\theta-\theta_2)\dots S^{c_N,b_N}_{c_{N-1},a_N}(\theta-\theta_2)\Bigr),
		\end{eqnarray}
		where the charge conjugation matrix for sine-Gordon model reads $C_{a,b} = \delta_{a+b,0}$. We note that if there exist bound states, the form factors will possess poles in addition. 
	\end{itemize}

	\section{Form factors of semi-local operators in the sinh-Gordon model}
	\label{FFShGM}
In this appendix we study the twist operator $\tau$ in the sinh-Gordon model. Its defining property is that it links the different symmetry sectors associated with the $\mathbb{Z}_2$ symmetry $\phi\to-\phi$. Thus the twist field is semi-local and can be viewed as analog of the order and disorder operators in the Ising field theory. Its form factors have been already introduced in~\cite{Horvath-20} in the context of symmetry resolved entanglement of integrable quantum field theory.

	\subsection{Two-particle form factors}
	\label{Two-particle Form Factors for semi-local operators in Sinh-Gordon model}
Following Ref.~\cite{KoubekMussardo93} we apply the form factor axioms to the twist operator. We start from deriving the two-particle form factor and then we are able to retrieve the general form for the allowed cases. This is equivalent to spanning the space of semi-local operators for the sinh-Gordon model. It is useful to introduce the minimal form factor, derived in Ref.~\cite{Fring-93}, with the properties 
	\begin{eqnarray}
		\label{MinimalFF}
		&& F_\mathrm{min}(\theta) = S(\theta)F_\mathrm{min}(-\theta),\\
		\label{GSemi-Locality}
		&& F_\mathrm{min}(\ii\pi-\theta) = F_\mathrm{min}(\ii\pi + \theta),\\
		&& F_\mathrm{min}(\theta)F_\mathrm{min}(\ii\pi+\theta) =  \frac{\sinh\theta}{\sinh\theta + \ii\sin\frac{\pi B}{2}},\label{FProp}
	\end{eqnarray}
	where the scattering matrix is given by \eqref{eq:ShGMSmatrix}, and $B=g^2/(8\pi+g^2)$ parametrises the interaction. Explicit expressions for the minimal form factor have been derived~\cite{Fring-93}. Using the minimal form factor, we make the ansatz 
	\begin{equation}
		\label{2PFF}
		F_2^{\mathcal{\tau}}(\theta_1,\theta_2) = F_0^\mathcal{\tau} H_2\,\frac{F_\mathrm{min}(\theta_1-\theta_2)}{\cosh\frac{\theta_1-\theta_2}{2}}
	\end{equation}
	for the two-particle form factor of the twist operator $\tau$, and we already used that its Lorentz spin vanishes~\cite{CardyMussardo90}. Now checking the form factor axioms we deduce that the scattering axiom is satisfied by virtue of (\ref{MinimalFF}) and the annihilation pole axiom requires $H_2=1/F_\mathrm{min}(\ii\pi)$ and $F_0^\tau$ being the expectation value of $\tau$ between the vacua in the different symmetry sectors. Furthermore, taking $\theta_1\to\theta_1+2\pi\ii$ we deduce 
		\begin{equation}
			\label{SLFactor}
			l(\mathcal{\tau}) = -1,
		\end{equation}
	reflecting the semi-locality of the twist operator. Hence, in total we find for the two-particle form factor 
	\begin{equation}
		\label{2PFF}
		F_2^{\mathcal{\tau}}(\theta_1,\theta_2) = \frac{F_0^\mathcal{\tau}}{F_\mathrm{min}(\ii\pi)}\frac{F_\mathrm{min}(\theta_1-\theta_2)}{\cosh\frac{\theta_1-\theta_2}{2}}.
	\end{equation}

	\subsection{General form factors}
	\label{General solution}
	To fix the general solution of the form factor axioms we assume the following parametrisation~\cite{Fring-93}
	\begin{equation}
		\label{FFShGparam}
		F^{\tau}_{N}(\theta_1,\theta_2,\dots,\theta_N) = H_NQ_N(\theta_1,\dots,\theta_N)\prod_{i<j}^{N}\frac{F_\mathrm{min}(\theta_i-\theta_j)}{\cosh{\frac{\theta_i-\theta_j}{2}}}
	\end{equation}
	with $Q_N$ a symmetric function of its arguments. It is straightforward to verify the analyticity, scattering and Lorentz axioms. For the periodicity axiom we get 
	\begin{eqnarray}
		&F^{\tau}_{N}(\theta_1 + 2\ii\pi,\theta_2,\dots,\theta_N)   \nonumber\\
		&= H_NQ_N(\theta_1,\dots,\theta_N)\prod_{2<i<j}^{N}\frac{F_\mathrm{min}(\theta_i-\theta_j)}{\cosh{\frac{\theta_i-\theta_j}{2}}}\prod_{k=2}^{N}\frac{F_\mathrm{min}(\theta_1-\theta_k+2\ii\pi)}{\cosh{\frac{\theta_1-\theta_k+2\ii\pi}{2}}} \nonumber\\
		& = (-1)^{N-1} H_NQ_N(\theta_1,\dots,\theta_N)\prod_{i<j}^{N}\frac{F_\mathrm{min}(\theta_i-\theta_j)}{\cosh{\frac{\theta_i-\theta_j}{2}}}
	\end{eqnarray}
	which confirms the semi-locality $l(\tau)=-1$ for even-particle form factors ($N$ even). Furthermore, the $Q_N$ have to be invariant by a shift of any individual rapidity by $2\pi\ii$, ie, $Q_N(\theta_1+2\pi\ii,\dots,\theta_N)=Q_N(\theta_1,\dots,\theta_N)$. The annihilation pole axiom takes the form of a recursion equation for the $Q_N$'s, namely
	\begin{eqnarray}
		\label{RecEq}
		&Q_{N+2}(\theta+\ii\pi,\theta,\theta_1,\dots,\theta_N)\nonumber\\ 
		&=\left[\prod_{k=1}^N(\sinh{(\theta-\theta_k)} + \ii\sin{\pi B}) + \prod_{k=1}^N(\sinh{(\theta-\theta_k)} - \ii\sin{\pi B}) \right]\nonumber\\*
		&\qquad\times Q_{N}(\theta_1,\dots,\theta_N).
	\end{eqnarray}
	It also affects normalisation $H_N$,
	\begin{equation}
		H_{N} = H_2\left(\frac{4\sin(\pi B/2)}{F_{min}(\ii\pi)}\right)^{2N-1}.
	\end{equation}
	
	\subsection{Vacuum expectation value}
	\label{Computing bartau}
	We would like to study the expectation value of the twist operator between the vacua in the sectors with periodic/antiperiodic boundary conditions, and in particular want to show that it is non-vanishing. Normalisation of any state in a finite-volume system implies
	\begin{equation}
		1 = \mbox{ }_\mathrm{NS}\langle 0|0\rangle_\mathrm{NS} = \mbox{ }_\mathrm{R}\langle 0| \tau(L)\tau(0)|0\rangle_\mathrm{R},
	\end{equation}
         which, by inserting the resolution of unity between the two twist operators, leads to
	\begin{equation}
		\label{TauBar}
		1 = (F_0^\tau)^2\sum_{N=0}^\infty\int\prod_{i=1}^{N}\frac{\dd p_i}{2\pi} |f_N^\tau(p_1,\dots,p_N)|^2 e^{\ii L\sum_{j=1}^{N}p_j},
	\end{equation}
	where we introduced 
	\begin{equation}
		F_N^\tau(p_1,\dots,p_N) = F_0^\tau\,f_N^\tau(p_1,\dots,p_N)
	\end{equation}
	and took the integrations over the individual momenta $p_i=M\sinh\theta_i$. It is then straightforward to check that the right-hand side of (\ref{TauBar}) is independent from $L$ since $p_jL=2\pi N_j$. We get the same result when anti-periodic boundary conditions are considered, since the ground state as well would transform under the action of translation operator and cancel an extra contribution in the exponent. There is no dependence on boundary conditions and infinite volume limit $L\to \infty$ can then be safely taken; yielding the result 
	\begin{equation}
		\label{TauBar2}
		F_0^\tau = \left[\sum_{N=0}^\infty\int\prod_{i=1}^{N}\frac{\dd p_i}{2\pi} |f_N^\tau(p_1,\dots,p_N)|^2\right]^{-1/2}.
	\end{equation}
	Given that each term in the sum is non-negative and fastly convergent~\cite{Mussardo10}, we conclude that $F_0^\tau$ is finite and non-vanishing.
		
	\subsection{Scaling dimension}
	\label{Scaling dimension}
	In order to compute the scaling dimension and identify the corresponding operator in the conformal field theory, we apply the $\Delta$-theorem~\cite{Delfino:96}. For simplicity we consider the non-interacting case $g=0$, where the trace of the stress-energy tensor $\Theta$ has only a non-vanishing form factor in the two-particle case. Thus we find
	\begin{eqnarray}
		\label{DeltaTh}
		\Delta_\tau  &=& -\frac{1}{4\pi^2 F_0^\tau}\int\dd^2\vec{r} \langle\Theta(0)\tau(\vec{r})\rangle \nonumber\\
		& =& - \frac{m^2}{2\pi F_0^\tau}\int_{0}^{\infty}\dd r\,r \int_{0}^{\infty}\frac{\dd\theta_1}{2\pi}\frac{\dd\theta_2}{2\pi} F_2^\tau(\theta_1,\theta_2)e^{-mr(\cosh\theta_1 + \cosh\theta_2)}\nonumber \\
		& =& - \frac{1}{2\pi}\int_{0}^{\infty}\dd x\,x \int_{-\infty}^{\infty}\frac{\dd\theta_1}{2\pi}\frac{\dd\theta_2}{2\pi} \frac{e^{- x(\cosh\theta_1 + \cosh\theta_2)}}{\cosh\frac{\theta_1-\theta_2}{2}},
	\end{eqnarray}
	where we used $F_\mathrm{min}(\theta)=1$ for $g=0$. The further analysis of (\ref{DeltaTh}) proceeds with
	\begin{enumerate}
		\item make use of the Werner formula $\cosh\theta_1 + \cosh\theta_2 = 2\cosh\frac{\theta_1+\theta_2}{2}\cosh\frac{\theta_1-\theta_2}{2}$ and perform the change of variables $\theta_+=\frac{\theta_1+\theta_2}{2}$, $\theta_-=\frac{\theta_1-\theta_2}{2}$ in the integral;
		\item exploit the integral representation~\cite{GradshteynRyzhik80} of the modified Bessel function $K_0(2x\cosh\theta_-) = \int_{0}^{\infty}\frac{\dd\theta_+}{2\pi} e^{-2x\cosh\theta_+\cosh\theta_-}$;
		\item exchange integration order between $x$ and $\theta_-$ and use the property of indefinite integrals $\int \dd x\,x K_0(x) = x K_1(x)$ to perform the integral (use 5.56(2) in Ref.~\cite{GradshteynRyzhik80});
		\item evaluate the remaining integral $\int_0^{\infty} \frac{\dd\theta}{2\pi}\frac{1}{4\cosh^3\theta}=\frac{1}{16}$.
	\end{enumerate}
	Thus we obtain the same scaling dimension as that of the twist operator $\tau$ of a massless free bosonic field theory~\cite{Ginsparg:notes}, which means that we can rightfully identify the semi-local operator of the sinh-Gordon theory with the former.
	
	\section{Initial states and pair structure}
	\label{Initial states and pair structure}
	We briefly give some additional motivation~\cite{FiorettoMussardo10,Sotiriadis-12} on our choice of the initial state \eqref{InitialState}, in particular its pair structure. First we note that for global quenches the initial state will be translationally invariant. Thus if expanded in the post-quench basis the initial state has the form
	\begin{equation}
		\label{key}
		|\psi\rangle = \sum_{N=0}^{\infty}\int_{-\infty}^{\infty}\prod_{k=1}^{N}\frac{\dd\theta_k}{2\pi}a_N(\theta_1,\dots,\theta_N)\, \delta\!\left(m\sum_{k=1}^N\sinh\theta_k\right) |\theta_1,\dots,\theta_N\rangle
	\end{equation}
	with some expansion coefficients $a_N(\theta_1,\dots,\theta_N)$. The zero-momentum condition is implemented by the Dirac delta function with the individual momenta $p_k=m\sinh\theta_k$. For simplicity we consider one particle species only. 

	In addition, for a global quench the initial state must be extensive in the sense that~\cite{CauxEssler13} 
	\begin{equation}
		\label{ExtInSt}
		\langle\psi|\psi\rangle \sim e^{LF(R)}, \quad L\to\infty,
	\end{equation}
	where $F$ depends on the characteristics of the quench and the regularisation parameter (such as the extrapolation time)  $R$. Obviously, the requirement \eqref{ExtInSt} implies that the state \eqref{key} indeed has to contain contributions with arbitrarily large particle numbers. Furthermore, as long as the expansion coefficients $a_N(\theta_1,\dots,\theta_N)$ are smooth functions the scaling of $\langle\psi|\psi\rangle$ will be $\sim L$ from the regularisation of the squared Dirac delta. Arguably the simplest way to overcome this is to consider initial states with the pair structure (\ref{InitialState}), which contain additional delta functions and thus lead to an exponential scaling in the system size. In this sense coherent squeezed states provide minimal solutions to the required conditions.
	
	\section{Finite-volume theory}
	\label{Finite volume theory}
	We regularise the theory using a finite volume, in particular it provides a regularisation for annihilation poles and the correct representation for the disconnected parts, as outlined in~\cite{DiSalvo:23, PozsgayTakacs08-1,PozsgayTakacs08-2}. 	Application of the Bethe ansatz takes care of defining a basis of energy eigenstates starting from the Bethe--Yang equations
	\begin{eqnarray}
		\label{BAE}
		&m_aL\sinh{\theta_j} = 2\pi I_j + \sum_{b}\sum_{k=1, k\neq j}^N\chi_{ab}{(\theta_k - \theta_j)}, \\ &\chi_{ab}(\theta) = -\ii\ln{S_{ab}(\theta)} , \quad \varphi_{ab} = \frac{\dd\chi_{ab}(\theta)}{\dd\theta},
	\end{eqnarray}
	where $I_j\equiv I_{j;a}$ implicitly contains the particle type $a$ which we suppress in the following, and $S_{ab}(\theta) = S_{ab}^{ba}(\theta)$ is the scattering matrix for the exchange between two-particles without production of particles of other species. In particular, these equations provide the link between the Bethe numbers $I_j$ and the corresponding rapidities $\theta_j$. Furthermore, we note that the scattering data of the theory enter via the kernels $\chi_{ab}$. We introduce the following quantities
	\begin{eqnarray}
		\label{FVDens}
		&\mathcal{J}_{j,l;a}(\theta_1,\dots,\theta_N;L) = 2\pi\partial_j I_{l;a} = m_a\,\delta_{j,l}\,\cosh{\theta_l} + \sum_b\varphi_{ab}(\theta_j - \theta_l), \\
		&\rho_{N;a}(\theta_1,\dots,\theta_N;L) = \mbox{det}\mathcal{J}_{j,l;a}(\theta_1,\dots,\theta_N;L).
	\end{eqnarray}
	As we are not interested in the full energy spectrum at finite volume, we only need few results on the Jacobian matrix and its determinant generated by the mapping (\ref{BAE}). We are then able to properly characterise any finite-volume state $|I_1,\dots,I_N\rangle_L$; the initial states we discussed in \ref{Initial states and pair structure} or the boundary initial state for instance (\ref{InitialState}) are then properly defined in the finite volume with the help of Bethe ansatz formalism~\cite{KormosPozsgay10} as follows
	\begin{eqnarray}
		\label{FVInitialSt}
		|\psi\rangle_L &=& \sum_{N=0}^{\infty}\sum_{0<I_1<\dots<I_N}\mathcal{N}_{2N;a}(\theta_1,\dots,\theta_N;L)\mathcal{K}_{2N;a}^b(\theta_1,\dots,\theta_N)\nonumber\\*
		& &\qquad\qquad\qquad\times |-I_N,I_N,\dots,-I_1,I_1\rangle_{b_1,\dots,b_N,L}.
	\end{eqnarray}
	Here we introduced short-hand notations for the normalisation and pair amplitude,
	\begin{eqnarray}
		\label{NDef}
		\mathcal{N}_{2N;a}(\theta_1,\dots,\theta_N;L)&=&\frac{\sqrt{\rho_{2N;a}(-\theta_N,\theta_N,\dots,-\theta_1,\theta_1;L)}}{\rho_{N;a}(\theta_1,\dots,\theta_N;L)},\\
		\label{KDef}
		\mathcal{K}_{2N;a}^b(\theta_1,\dots,\theta_N)&=&K_{a_1}^{b_1}(\theta_1)\dots K_{a_N}^{b_N}(\theta_N),
	\end{eqnarray}
	with the amplitude for individual pairs $K(\theta)$. Moreover, in large but finite volume, the $\rho_N$ functions (which can be regarded as the Jacobian of the change of basis) have a simple form up to small corrections,
	\begin{equation}
		\label{LVDens}
		\rho_{N;a}(\theta_1,\dots,\theta_N;L) = (m_aL)^N\prod_{k=1}^N\cosh{\theta_k}\left[1+O\left(\frac{1}{L}\right)\right].
	\end{equation}
	Thus the dominant contributions in the determinant are given by the diagonal ones. Furthermore, the volume dependence in the normalisation (\ref{NDef}) drops out.
	
	When the representation (\ref{FVInitialSt}) is employed in order to compute expectation values in the large-volume regime, one subtlety must be considered: since the discrete sums are turned into integrals, the restriction over identical particles (given by the fermionic nature of the particles themselves) is loosened, which means that those states (ie, those which contain two equal particles) also contribute to the final result. In principle we would need to subtract these contributions any time one considers integrals that overlap those regions; however they only show up in orders higher in $K$ than those we consider.
	
	Matrix elements are mapped into the new basis counterparts via (\ref{FVDens}), up to corrections $\mathcal{O}(e^{-mL})$, ie,
	\begin{equation}
		\label{FVFF}
		\hspace{-10mm}\langle I_1,\dots,I_N|O|J_1,\dots,J_M\rangle_L\! = \!\frac{F^{O}_{N+M}(\xi_1+\ii\pi,\dots,\xi_N+\ii\pi,\theta_1,\dots,\theta_M)}{\sqrt{\rho_N(\xi_1,\dots,\xi_N;L)}\sqrt{\rho_M(\theta_1,\dots,\theta_M;L)}}.
	\end{equation}
	The normalisation of the basis states is chosen to be
	\begin{equation}
		\label{FVStNorm}
		\langle I_1,\dots,I_N|J_1,\dots,J_M\rangle_L = \delta_{N,M}\delta_{I_1,J_1}\dots\delta_{I_N,J_M}.
	\end{equation}
	We also briefly recall here the main formula for squared delta function,
	\begin{eqnarray}
		\label{DeltaReg}
		\delta^2(\xi - \theta) = mL\cosh\theta\,\delta(\xi-\theta) + O\left(\frac{1}{L}\right)\,
	\end{eqnarray}
	which is needed to regularise the norm of initial states and disconnected parts; the latter are contained in diagonal matrix elements and can be resolved with the following
	\begin{eqnarray}
		\label{FVDiagElem}
		&&\bra{I_1,\dots,I_N}\mathcal{O}\ket{I_1,\dots,I_N}_{a_1,\dots,a_N} = \frac{1}{\rho_{N;a}(\theta_1,\dots,\theta_N;L)}\nonumber\\*
		&&\qquad\times\sum_{A\subset\{1,\dots,N\}}\rho_{N;a}(\theta_1,\dots,\theta_N|A;L)F^\mathcal{O}_{2|A|,\mathrm{conn}}(\{\theta_k\}_{k\in A}).
	\end{eqnarray}
	This was introduced in~\cite{PozsgayTakacs08-2} and valid for local operators; the set $A$ labels any combination of rapidities $\{\theta_k\}_{k\in A}$ that contributes to the connected part of the form factor~\cite{LeclairMussardo99} $F^\mathcal{O}_{2|A|,\mathrm{conn}}$, while the remainder $\theta_1,\dots,\theta_N|A$ contributes as a Kronecker delta. This formula provides as well a regularised version of possible divergent parts contained in its left-hand side in terms of connected form factors. Usually~\cite{PozsgayTakacs08-1}, this is the only case when annihilation poles are relevant in finite volume too, due to rapidity shift induced by Bethe ansatz equations (\ref{BAE}); however, the pair structure makes it possible to have two identical rapidities in ingoing and outgoing states, requiring to slightly generalise~\cite{PozsgayTakacs08-2} the previous expression to the case when two states contain different Bethe numbers.
	
	Both of these two results can be derived for a generic integrable system, but as pointed out in the case of Ising~\cite{Calabrese-12jsm1,SE12} and sine-Gordon models~\cite{BSE14} for semi-local operators, one should always take care in avoiding the breaking of a discrete symmetry of the ground state, be it the fermionic parity or the topological equivalence of the vacua. In a generic case we can relate this $\mathbb{Z}_2$-symmetry to the transformation $\mathcal{T}$ and call the two sectors Ramond (R) and Neveu--Schwarz (NS). These sectors can be formally related to the presence of periodic or anti-periodic boundary conditions, as assumed in Section \ref{Local and semi-local operators}. Then we can formally express the ground state as
	\begin{equation}
		\label{SLSplitting}
		|0\rangle_\mathrm{R,NS} = \frac{|0\rangle_+\pm|0\rangle_-}{\sqrt{2}},
	\end{equation}
	where the subindices $\pm$ stand for the eigenvalues of the transformation $\mathcal{T}$. There are of course cases when the underlying global discrete symmetry is not simply $\mathbb{Z}_2$; in those, semi-local indices are not just $\pm1$ and more sectors are allowed; in other words, when it is needed to be studied an operator that interpolates between more exotic sectors, this splitting must be further considered in detail. In some sense all the operators we are studying are different representations of the same twist operator between Ramond and Neveu--Schwarz sectors.
	
	When local operators $\mathcal{O}$ are considered, splitting between the two sectors is tantamount, since local operators have zero expectation value between the sectors. Thus for any initial boundary state $|\psi\rangle$ and summing over the eigenvalues $|k\rangle_a$ in the relative sector we obtain
	\begin{eqnarray}
		\label{LOS}
		\langle\psi|\mathcal{O}|\psi\rangle_L &=& \frac{1}{2}\left[{}_\mathrm{R}\langle\psi|\mathcal{O}|\psi\rangle_\mathrm{R} + {}_\mathrm{NS}\langle\psi|\mathcal{O}|\psi\rangle_\mathrm{NS}\right] \nonumber\\
		&=&\frac{1}{2}\sum_{a=\mathrm{R,NS}} \sum_{|k\rangle_a}{}_a\langle k|\mathcal{O}|k\rangle_a\,\Big|\,_a\langle k|\psi\rangle_a\Big|^2.
	\end{eqnarray}
	Thus one can rewrite the two sums as one over the complete Hilbert space without any loss of information. In principle one can also sum over one of the two sectors and, taking the thermodynamic limit, retrieve the same result; this can also be guessed from the fact that (anti-)periodic boundary conditions imply the sum over (half-)integer Bethe numbers.
	
	In the case of semi-local operators the splitting has a non-trivial impact on the computation of expectation values: following Sec.~\ref{Global quantum quenches in Integrable Field Theory} they must be evaluated between two different sectors, hence the expectation value after the quench vanishes at late times is given by
	\begin{equation}
		\label{NLOS}
		\langle\psi|\tilde{\mathcal{O}}|\psi\rangle_L = \,_\mathrm{NS}\langle\psi|\tilde{\mathcal{O}}|\psi\rangle_\mathrm{R}.
	\end{equation}
	We are going to relate this to the presence/absence of polynomial tails in time in the next sections by the means of linked cluster expansion.

	\section{Exponential decay from annihilation poles}
	\label{Exponential decay from annihilation poles}
	In this section we show how exponential decay is obtained in the framework of the quench action method, where the derivation is much simpler and clearer as compared to the linked clusted expansion.
	It requires in fact the following steps:
	\begin{enumerate}
		\item Compute the overlap $\langle\rho|\psi\rangle$ in (\ref{QAm}), which is the denominator of the expression. It is of the order $\mathcal{O}(L^N)$ where $N$ is the number of particle pairs in the finite-volume representation.
		\item Write down the numerator: time dependence is only contained in the exponential oscillations, but the analytic structure of the form factors contains in the connected part double poles for the annihilation pole axiom; when one of them is picked, the derivative over the rapidity gives a time contribution. We shall consider other contributions from the residue of the double pole as regular terms without any polynomial prefactor depending on $t$ and we do not regard them for the remainder of this analysis.
		\item Then, in the sum over the number of particle pairs contained in the initial state $M$, find the lowest order in $L$ state that allows to extract the $N$ double poles contained in the form factors; this term will not be vanishing in the thermodynamic limit. When there is more than one state that respects this condition, consider the sum of the different contributions.
		\item At this stage we can explicitly pick the poles by using the following formula for extracting dominating contributions in the selected matrix elements,
		\begin{eqnarray}
			\label{HOFF}
			&\langle\xi_1,-\xi_1,\dots,\xi_N,-\xi_N|\mathcal{O}|-\rho_M,\rho_M,\dots,-\rho_1,\rho_1\rangle\\
			&\qquad\sim \sum_{a_1,\dots,a_p=1}^{|N-M|}F_{2|N-M|}^{\mathcal{O}}(\xi_{a_1},-\xi_{a_1},\dots,\xi_{a_p},-\xi_{a_p})\nonumber\\*
			&\qquad\qquad\times\prod_{b=1}^{\min(N,M)} \left[\frac{4i}{\left(\xi_{\sigma_a(b)} - \rho_b\right)^2} - \delta^2(\xi_{\sigma_a(b)} - \rho_b)\right],\nonumber
		\end{eqnarray}
		where $p=|N-M|$. Afterward we perform the ratio between nominator and denominator.
		\item Then we extract the double poles and we apply the regularised squared Dirac deltas and the thermodynamic limit can be taken, bringing up the exponential decay with 
		\begin{eqnarray}
			\label{ExpDecayQAm}
			\prod_{b=1}^{\min(N,M)} \left[1 - 4\frac{t}{L}\tanh\rho_b \right] &= \exp\left\{\sum_{b=1}^{\min(N,M)}\log{\left[1 - 4\frac{t}{L}\tanh\rho_b \right] }\right\}\nonumber \\
			&\simeq e^{-\Gamma t},
		\end{eqnarray}
		where the decay rate is
		\begin{equation}
			\label{FullGamma}
			\Gamma = \frac{2m}{\pi}\int_{0}^{\infty}\dd\theta\rho(\theta)\sinh\theta,
		\end{equation}
		using the Yang--Yang approach to thermal equilibrium physics to turn the sum into an integral weighted by the density of states $L\rho(\theta)$.
		In the limit of small quench parameter $K(\theta)$, where the particle density in the steady state can be analytically found and we retrieve (\ref{Gamma}).
	\end{enumerate}
	The prefactor of the previous expressions shows up in the general formula (\ref{Main}) as the function $f(t)$. It is in the end straightforward linking the presence of exponential decay to the analytic structure underlying form factors and operators' matrix elements in general, given (\ref{HOFF}) and its consequences.

	\section{Exponential decay and power-law behaviour from thermal form factors}
	\label{Exponential decay and power law behaviour from thermal Form Factors}
	In this section we introduce and employ the formalism of thermal form factors to recover features we described using linked cluster expansion and quench action method. Working in the quench action framework, we can study the matrix element of the operator $\mathcal{O}$ (may it be local or semi-local) between the initial state $|\psi\rangle$ and the steady one $|\rho\rangle$ (\ref{QAm}). To the extent of reproducing the result (\ref{Main}), the key idea lies in splitting the matrix elements in the numerator of the right-hand side as a sum between their regular and the singular part,
	\begin{eqnarray}
		\label{MEsep}
		\langle\xi_1,\dots,\xi_M|\mathcal{O}|\theta_N,\dots,\theta_1\rangle &= \langle\xi_1,\dots,\xi_M|\mathcal{O}|\theta_N,\dots,\theta_1\rangle_\mathrm{sing} \nonumber\\
		& \quad + \langle\xi_1,\dots,\xi_M|\mathcal{O}|\theta_N,\dots,\theta_1\rangle_\mathrm{reg} .
	\end{eqnarray} 
	This is always allowed for integrable theories since we know that form factors are meromorphic functions in the physical strip $0<\mbox{Im}(\theta)<\pi$; hence we can write in any point their Laurent expansion and isolate the singular contributions. The regular contribution in (\ref{MEsep}) can be described by the means of the following prescription
	\begin{eqnarray}
		\label{RegP}
		&\langle\xi_1,\dots,\xi_M|\mathcal{O}|\theta_N,\dots,\theta_1\rangle_\mathrm{reg}  \nonumber\\
		&\quad=\lim_{\{\epsilon_i\}\to 0^+}  \mathcal{FP}\left\{\langle\xi_1+\ii\epsilon_1,\dots,\xi_M+\ii\epsilon_M|\mathcal{O}|\theta_N,\dots,\theta_1\rangle \right\}.
	\end{eqnarray}
	Here $\mathcal{FP}$ denotes the finite part as introduced in (\ref{ConnFormFactors}), it is equivalent to removing singular contributions in its argument. Extending the same notation to form factors, we get
	\begin{eqnarray}
		\label{RegFF}
		&\hspace{-15mm}F^\mathcal{O}_{N+M,\mathrm{reg} }(\xi_1+\ii\pi,\dots,\xi_M+\ii\pi,\theta_N,\dots,\theta_1) \nonumber\\
		&\hspace{-10mm} =\lim_{\{\epsilon_i\}\to 0^+}  \mathcal{FP}\left\{F^\mathcal{O}_{N+M }(\xi_1+\ii\pi+\ii\epsilon_1,\dots,\xi_M+\ii\pi+\ii\epsilon_M,\theta_N,\dots,\theta_1) \right\},
	\end{eqnarray}
	the same prescription as taking the Laurent expansion for $\{\epsilon_i\}\to 0^+$ and subtracting the pole contributions. 	These are the only ones that have to be removed since form factors are meromorphic functions. The singular part is responsible for the exponential relaxation of the observable towards the steady state expectation values (see \ref{Exponential decay from annihilation poles}), while the regular part at the thermodynamic limit should tend to zero when expectation values over the post-quench steady state are not considered~\cite{Essler24}. Both these results can be retrieved by means of thermal form factors as we show in this Appendix.
	
	We shall employ again the finite-volume scheme (see \ref{Finite volume theory} for more details) in order to regularise divergent parts of the expressions. In this case we need to choose a particular representative state $|\rho\rangle_L$ to take into account of symmetry breaking in finite volume that is still an eigenvalue of the post-quench Hamiltonian
	\begin{equation}
		H|\rho\rangle_\mathrm{R} = E_{\rho,\mathrm{R}}|\rho\rangle_\mathrm{R}.
	\end{equation}
	We are free to choose any of the two sectors and picked the Ramond sector here. We can then compute the right-hand side of (\ref{QAm}) in the finite volume; now the sum is carried over one Hilbert space (related to the initial state) instead if two as before. We start by splitting (\ref{QAm}) into regular and singular parts. Considering only the former we project the initial state onto the basis $|\phi\rangle$,
	\begin{eqnarray}
		\label{RHSreg}
		&\frac{\langle\psi|\mathcal{O}(t)|\rho\rangle_\mathrm{reg}}{\langle\psi|\rho\rangle} \nonumber\\ 
		&\qquad =\sum_{|\phi\rangle}e^{\mathcal{E}_\phi-\mathcal{E}_\rho+i(\omega_\phi-\omega_\rho)t}{}_\mathrm{R,NS}\langle\xi_1,\dots,\xi_{M_\phi}|\mathcal{O}|\rho_N,\dots,\rho_1\rangle_\mathrm{R,reg},
	\end{eqnarray}
	where
	\begin{eqnarray}
		&|\phi\rangle = |I_1,\dots,I_{M_\phi}\rangle_\mathrm{R,NS} = |\xi_1,\dots,\xi_{M_\phi}\rangle_\mathrm{R,NS},\nonumber \\
		&\mathcal{E}_\phi = \log\langle\psi|\phi\rangle, \qquad 
		\omega_\phi = m\sum_{i=1}^{M_\phi}\cosh\xi_i.\label{Overlap}
	\end{eqnarray}
	Terms with $M_\phi<N$ are negligible contributions in the thermodynamic limit. We shall not consider them since they are not contributing to the expectation value.
	
	By the means of saddle-point approximation, we can now estimate (\ref{RHSreg}) in the late-time regime since the integrand in the large-volume approximation is a smooth function. We can then show that these contributions have polynomial tails at late times that dominate the exponential decay given by the annihilation poles, as we expected. Also, we can see that semi-local operators do not allow polynomial contributions, namely either the overlap $e^{\mathcal{E}_\phi}$ vanishes when $|\phi\rangle\in \mathrm{NS}$ or the matrix element when $|\phi\rangle\in \mathrm{R}$, as expected from (\ref{SemiLocalME}).
	
	We note, however, that the expansion in excitations over the representative state is not well controlled, and one has to  carefully consider the scaling of such matrix elements in the thermodynamic limit. In fact, it has been shown for the Lieb--Liniger model~\cite{Essler24} that only few (rare) matrix elements from the infinitely many possible ones scale in such a way as to lead to finite results in (\ref{RHSreg}). A possible way to organise the expansion is given using $1/c$ as control parameter~\cite{GranetEssler21}, with $c$ denoting the interaction strength in the Lieb--Liniger model. Also the subset of relevant matrix elements may depend on the considered operator, eg, for field powers in the sinh-Gordon model one has to include different particle excitations instead of particle-hole pairs~\cite{DiSalvo:23}. For what regards the results presented here, we have checked each of them that the small quench limit agrees with linked cluster expansion.
	
	At last, we would like to check that the small quench regime of the dominating polynomial contributions agree with the result in the linked cluster expansion. In order to do so, we need to introduce excitations over the representative state (particles and holes) and to compute the matrix elements of local operators of such species, ie, the thermal form factors.

	\subsection{Thermal form factor axioms}
	\label{Thermal Form Factor axioms}
	In this section we review the concept of thermal form factors as originally introduced by Cort\'{e}s Cubero and Panfil~\cite{CuberoPanfil19,CuberoPanfil20} and further developed in~\cite{PanfilKonik23}. Here we consider local operators $\mathcal{O}$ only, the generalisation will be discussed in \ref{Exponential decay} below. 
	
	The thermal form factors are matrix elements of some operator over the thermal (or more general multi-particle) state and excitations on top of that. We define 
	\begin{eqnarray}
		\label{TFF:Def}
		F^\mathcal{O}_\rho(\theta_1,\dots,\theta_N)& = \langle\rho|\mathcal{O}|\rho,\theta_1,\dots,\theta_N\rangle \\&= \lim_{L\to\infty}\sqrt{\frac{\rho_L(\{\rho\},\theta_1,\dots,\theta_N)}{\rho_L(\{\rho\})}}\langle\{I\}|\mathcal{O}|\{\bar{I}\},I_1,\dots,I_N\rangle,\nonumber
	\end{eqnarray}
	where $\ket{\rho}$ denotes the background state defined by its distribution of rapidities obtained from the Bethe numbers $\{I\}$, the rapidities $\theta_j$ obtained from the Bethe numbers $I_j$ are the excitations on top of this distribution, and $\{\bar{I}\}$ denotes the shifted Bethe numbers due to the presence of the excitations. For simplicity we consider only excitations without internal degrees of freedom, ie, the Ising and sinh-Gordon models, and without bound states.
	
	Similar to the form factors above the vacuum, the thermal form factors have to satisfy a set of axioms~\cite{CuberoPanfil19,CuberoPanfil20,PanfilKonik23}:
	\begin{enumerate}
		\item Scattering axiom: 
		\begin{eqnarray}
			\label{Sc:Ax}
			&F^\mathcal{O}_\rho(\theta_1,\dots,\theta_i,\theta_{i+1},\dots,\theta_N) \nonumber\\
			&\qquad= S(\theta_i-\theta_{i+1})F^\mathcal{O}_\rho(\theta_1,\dots,\theta_{i+1},\theta_{i},\dots,\theta_N).
		\end{eqnarray}
		\item Crossing symmetry: 
		\begin{equation}
			\label{Cross:Ax}
			\langle \rho,\xi |\mathcal{O}| \rho,\theta_1,\dots,\theta_N\rangle=F^{\mathcal{O}}_{\rho}(\xi +\ii\pi,\theta_1,\dots,\theta_N).
		\end{equation}
		\item Periodicity axiom: 
		\begin{equation}
			\label{Period:Ax}
			\hspace{-10mm}
			F^\mathcal{O}_{\rho}(\theta_1,\dots,\theta_N) = R(\theta_N|\theta_1,\dots,\theta_N)F^\mathcal{O}_{\rho}(\theta_N+2\ii\pi,\theta_1,\dots,\theta_{N-1}).
		\end{equation}
		We recall that the operator $\mathcal{O}$ is assumed to be local. The factor $R(\theta_N|\theta_1,\dots,\theta_N)$, which will be discussed below, takes into account the scattering of the excitations from the background.
		\item Annihilation pole axiom: 
		\begin{eqnarray}
			\label{AnnPol:Ax}
			&&\mbox{Res}\left[F^\mathcal{O}_{\rho}(\theta',\theta,\theta_1,\dots,\theta_N),\theta'=\theta+\ii\pi\right] \nonumber\\
			&&\quad = \ii\left(1-R(\theta|\theta_1,\dots,\theta_N)\prod^N_{k=1}S(\theta-\theta_k)\right)F_\rho^\mathcal{O}(\theta_1,\dots,\theta_N). 
		\end{eqnarray}
		The additional factor $R(\theta_N|\theta_1,\dots,\theta_N)$ is again due to the scattering of the particles in the background state. 
		\item Cluster properties: If one boosts the rapidities of a subset of the particles to an infinite distance, the form factors are expected to factorise as 
		\begin{eqnarray}
			\label{Cluster:Ax}
			\lim_{\Lambda\to\infty}&F^{\mathcal{O}}_{\rho}(\theta_1+\Lambda,\dots,\theta_M+\Lambda,\theta_{M+1},\dots,\theta_N)\\& = \mathcal{N}^{\mathcal{O}}F^{\mathcal{O}_1}(\theta_1,\dots,\theta_M)F^{\mathcal{O}_2}_\rho(\theta_{M+1},\dots,\theta_N)\nonumber
		\end{eqnarray}
		where the two operators $\mathcal{O}_{1,2}$ depend on the internal symmetries of the theory and the fusion rules of the related conformal field theory. Note that the first form factor on the right-hand side is defined over the vacuum.
	\end{enumerate}
	Semi-local operators $\tilde{\mathcal{O}}$ will be considered below, where the symmetry breaking in finite volume and the dependence on the choice of the representative state have to be treated carefully.
	
	Now let us further discuss the factor $R$. This function is defined starting from the backflow of the background state: when a new particle is introduced on top of the background state, the latter is modified in order to preserve the Bethe ansatz prescription. The single-particle backflow $F_\mathrm{F}(\tilde\theta|\theta_1)$ is the quantity we are interested in, in order to build the $R$ function for a fermionic system, and it can be expressed with a TBA-like equation
	\begin{equation}
		\label{BackFlowF:eq}
		2\pi F_\mathrm{F}(\tilde\theta|\theta_1) = \delta_\mathrm{F}(\tilde\theta-\theta_1) + \int_{-\infty}^{\infty}\frac{\dd\theta'}{2\pi}\vartheta(\theta')\varphi(\tilde\theta - \theta')F_\mathrm{F}(\theta'|\theta_1),
	\end{equation}
	where $\vartheta$ is the filling fraction of the representative state $|\rho\rangle$, $\varphi$ the kernel of the TBA equation and $\delta_\mathrm{F}$ its primitive (ie, the logarithm of the S-matrix). The bosonic version is analogous. Then we can write the one-particle $R$ function as
	\begin{equation}
		\label{1pR:Def}
		R(\tilde{\theta}|\theta_1)=e^{\ii[2\pi F_F(\tilde\theta|\theta_1)-\delta(\tilde{\theta}-\theta_1)]},
	\end{equation}
	with the multiparticle version simply be given by the product,
	\begin{equation}
		\label{R:Def}
		R(\tilde\theta|\theta_1,\dots,\theta_N) = \prod^{N}_{j=1}R(\tilde{\theta}|\theta_j)
	\end{equation}
	from the elasticity and factorisability of scattering processes in integrable models.
	
	We are now able to construct the two-particle thermal form factors; these are useful in order to construct one particle/hole excitations over the representative state and to check their contributions to the polynomial tail in the time evolution of local operator expectation values. The workflow consists in defining the following functions, among which are the minimal form factors,
	\begin{eqnarray}
		\hspace{-10mm}C_\rho(\xi|\theta_1,\dots,\theta_N) = \frac{\xi}{\pi^2}\int_{-\infty}^{\infty}\dd\theta'\log R(\theta'|\theta_1,\dots,\theta_N)e^{\ii\frac{\xi\theta'}{\pi}},\\
		\hspace{-10mm}A(\theta;\theta_1,\dots,\theta_N) = \exp\left[-\frac{1}{4}\int_{-\infty}^{\infty}\frac{\dd\xi}{\xi}\frac{C(\xi)+C(\xi|\theta_1,\dots,\theta_N)}{\sinh\xi}e^{-\xi(\ii\theta/\pi+1)}\right], \\
		\hspace{-10mm}F^\mathrm{min}_\rho(\theta_1,\theta_2) \nonumber\\ 
		\qquad=e^{\frac{\pi}{4}\left[\frac{\partial C(\xi|\theta_1)}{\partial\xi}|_{\xi=0} + \frac{\partial C(\xi|\theta_2)}{\partial\xi}|_{\xi=0}\right]} A(\theta_1;\theta_2)A(\theta_2;\theta_1)F^\mathrm{min}(\theta_1-\theta_2).
	\end{eqnarray}
	Then the generic two-particle thermal form factor has the following form
	\begin{equation}
		\label{2pFF:Def}
		F^{\mathcal{O}}_\rho(\theta_1,\theta_2) = K^{\mathcal{O}}_\rho(\theta_1,\theta_2)F^\mathrm{min}_\rho(\theta_1,\theta_2) ,
	\end{equation}
	where the $K^\mathcal{O}_\rho$ ensures the corrext annihilation poles and clustering properties. It has to be stressed though that, since $\mathcal{O}$ is local, the two-particle form factor does not contain any annihilation pole.
	
	\subsection{Exponential decay}
	\label{Exponential decay}
	The same procedure we described in~\ref{Exponential decay from annihilation poles} can be applied (straightforwardly in the case of local operators, while when semi-local operators are studied the construction must be modified as described later) by expanding the initial state in  the space of excitations over the representative state. It turns out being useful when the overlap $\langle\rho|\psi\rangle_L$ is known in the thermodynamic limit~\cite{DeNardisCaux14,DeNardis-14,DeNardis-15,CauxEssler13,Caux16}. An assumption one can make is that
	\begin{equation}
		\label{QALocalAssump}
		\frac{\langle\rho|\mathcal{O}|\psi\rangle_{\mathrm{sing},L}}{\langle\rho|\psi\rangle_L} \simeq \sum_{\{\bold{p},\bold{h}\}}\langle\rho|\mathcal{O}|\rho',\{\bold{p},\bold{h}\}\rangle_L e^{-\delta s_{\bold{p},\bold{h}}},
	\end{equation}
	ie, the physically meaningful part of the initial state can be expanded in excitations over the representative state. Given that the operator $\mathcal{O}$ does not create an extensive number of excitations~\cite{CauxEssler13}, we can expand $\ket{\psi}$ as particle-hole pairs over the sea of excitations in the representative state $\ket{\rho}$. In case of large but finite volume $L$ the matrix elements can be approximated by the thermal form factors discussed above. Following~\cite{BSE14,PanfilKonik23} we can derive their most singular part using the annihilation-pole axiom as well as the periodicity axiom in the presence of a non-trivial background,
	\begin{eqnarray}
		\label{HOTFF}
		\bra{\rho}\mathcal{O}&\ket{\rho,\{\bold{p},\bold{h}\}}\sim\sum_{a_1,\dots,a_p=1}^{|N-M|}F_{\rho,2|N-M|}^{\mathcal{O}}(\xi_{a_1},-\xi_{a_1},\dots,\xi_{a_p},-\xi_{a_p})\nonumber\\*
		& \times\prod_{b=1}^{\min(N,M)} \left[\frac{4\ii}{\rho^\mathrm{tot}(\xi_{\sigma_a(b)})\left(\xi_{\sigma_a(b)} - \rho_b\right)^2} - \delta^2(\xi_{\sigma_a(b)} - \rho_b)\right],
	\end{eqnarray}
	a formula close to (\ref{HOFF}), except that the multiplicative form factor in front is now a thermal one, and the contribution from the total density of particles and holes with rapidity $\xi$. Hence, one gets a dressed exponential relaxation with decay rate
	\begin{equation}
		\label{FullGammaDr}
		\Gamma^\mathrm{dr} = \frac{2m}{\pi}\int_{0}^{\infty}\dd\theta\frac{\rho(\theta)}{\rho^\mathrm{tot}(\theta)}\sinh\theta.
	\end{equation}
	Also the prefactor $f(t)$ of the exponential has to be determined by resumming the contributions from the particle-hole pairs over the representative state. However, both prefactor and (\ref{FullGammaDr}) agree with previous expressions when the small quench limit is taken and the background reduces to the post-quench ground state.
	
	When semi-local operators are considered, the splitting between different sectors makes (\ref{QALocalAssump}) unpractical, since now we have 
	\begin{equation}
		\label{QASemiLocalAssump}
		\frac{\langle\rho|\tilde{\mathcal{O}}|\psi\rangle_L}{\langle\rho|\psi\rangle_L} = \frac{_\mathrm{R}\langle\rho|\tilde{\mathcal{O}}|\psi\rangle_\mathrm{NS}}{_\mathrm{R}\langle\rho|\psi\rangle_\mathrm{R}} ,
	\end{equation}
	and it is impossible to decompose the Neveu--Schwarz sector's initial state on the excitations over the steady state. Thus we have to work with the dual of the representative state in the Neveu--Schwarz sector $\ket{\rho}_\mathrm{NS}$ by the formal action of a twist operator $\sigma_{\mathbb{Z}_2}$~\cite{Doyon07-2}
	\begin{equation}
		\label{rhoNS}
		\ket{\rho}_\mathrm{NS} = \lim_{x\to 0^+}\sigma_{\mathbb{Z}_2}(x)\ket{\rho}_\mathrm{R}.
	\end{equation}
	Here we have introduced the position of the twist operator $x$, from which its branch cuts runs to $+\infty$. Furthermore, without loss of generality the operator $\tilde{\mathcal{O}}$ in \eqref{QASemiLocalAssump} is assumed to be located at the origin. The composite operator is defined by using a point-splitting prescription~\cite{Ginsparg:notes}, $\mathcal{O}'=\lim_{x\to 0^+} \tilde{\mathcal{O}}\,\sigma_{\mathbb{Z}_2}(x)$, by contruction it preserves the translation invariance of the system. Simple examples of a twist operator are just the magnetisation $\sigma$ or disorder $\mu$ in the Ising model. 

Hence the matrix element (\ref{QASemiLocalAssump}) can be cast into the same shape of the local case
	\begin{equation}
		\frac{_\mathrm{R}\langle\rho|\tilde{\mathcal{O}}|\psi\rangle_\mathrm{NS}}{_\mathrm{R}\langle\rho|\psi\rangle_\mathrm{R}}  \simeq \sum_{\{\bold{p},\bold{h}\}} {}_\mathrm{R}\langle\rho|\mathcal{O}'|\rho',\{\bold{p},\bold{h}\}\rangle_\mathrm{R} \,e^{-\delta s_{\bold{p},\bold{h}}}.
	\end{equation}
It is then straightforward to extract the same exponential behaviour as it was done for the local case. 

	\subsection{Power-law contributions: Ising and sinh-Gordon models}
	\label{Examples: Ising and Sinh-Gordon models}
	In this section we work out some examples for the computation of time time evolution of local operator $\mathcal{O}$ in the quench action formalism. For the Ising model we just need the results found in~\cite{LeClair:96, Doyon:05} and do not have apply the full thermodynamic bootstrap program outlined above. On the other hand, for interacting theories such as the sinh-Gordon model, we are going to need that.
	
	In the Ising model, we are going to study the energy operator $\epsilon$, which is a local relevant one. Its only non-vanishing form factors are 
	\begin{equation}
		F^\epsilon_\rho(\theta_1,\theta_2) = m\sinh\left(\frac{\theta_1-\theta_2}{2}\right),
	\end{equation}
	which are completely regular in the physical strip. We can insert them in the finite-volume expression (\ref{RHSreg}) and take the thermodynamic limit with the result
	\begin{equation}
		\langle\epsilon(t)\rangle_\psi  = \langle\epsilon(0)\rangle_\rho + \mathfrak{Re}\left\{\int_{0}^{\infty}\frac{\dd\theta}{2\pi}\frac{m\sinh\theta K(\theta)}{1+|K(\theta)|^2}e^{2\ii mt\cosh\theta}\right\},
	\end{equation}
	recovering the result by Fioretto and Mussardo~\cite{FiorettoMussardo10}. In the saddle point limit, we can see that the relevant contribution of the integral comes from the limit $\theta\to0$ and scales in time as $t^{-3/2}$, ie, a power law at late times. This example just shows the consistency of our approach compared to the known literature.
	
	In the sinh-Gordon model we consider as well a local relevant operator, namely the vertex operator $e^{\ii\alpha\phi}$. We do so for the following reasons: first of all, it leads to a very simple two-particle form factor (\ref{2pFF:Def}); second, we can generate all the other local fields expectation values by taking its power series (by correctly taking into account the operator mixing). Then the two-particle thermal form factor is just given by its vacuum counterpart multiplied by
		\begin{equation}
		\label{ShG2pTFF}
		\mathcal{N}_\rho^\alpha = \left[\frac{f_1^\alpha}{\mathcal{G}(\alpha)}\right]^2\langle e^{\ii\alpha\phi}\rangle_\rho,
	\end{equation}
	where the term in squared brackets does not depend on the representative state's density of states (or filling fraction). We formally know the full expression (\ref{2pFF:Def}) for the vertex operator; it depends only on the difference of the two rapidities and is regular in the physical strip. We can at this point describe the contribution to the time evolution of a single particle/hole excitation, which reads
	\begin{eqnarray}
		\label{1p/hContr}
		\langle e^{\ii\alpha\phi(t)}\rangle_\psi &= \langle e^{\ii\alpha\phi}\rangle_\rho + \frac{1}{2}\mathfrak{Re}\int_{0}^{\infty}\frac{\dd\theta}{2\pi}K(\theta)\vartheta^{\mathrm{(h)}}(\theta)F^{\alpha}_\rho(-\theta,\theta)e^{2\ii mt\cosh\theta} \nonumber\\
		&+ \frac{1}{2}\mathfrak{Re}\int_{0}^{\infty}\frac{\dd\theta}{2\pi}\frac{\vartheta^\mathrm{(p)}(\theta)}{K(\theta)}F^{\alpha}_\rho(-\theta,\theta)e^{2\ii mt\cosh\theta},
	\end{eqnarray}
	where $\vartheta^\mathrm{(h)}$ and $\vartheta^\mathrm{(p)}$ are the filling fractions for holes and particles respectively. 	The saddle point approximation, together with the scattering axiom (\ref{Sc:Ax}), ensures that the late-time behaviour is again $\sim t^{-3/2}$. 	One can see that the presence of interaction marks the difference between particle and hole contributions.
	
	\subsubsection{Small quench limit}
	\label{Small quench limit}
	The expression (\ref{1p/hContr}) can be compared to the linked cluster expansion by considering the small-quench limit $|K(\theta)|\ll1$. We are interested in the second order of the expansion. In this regime, we can firstly see the following relation
	\begin{equation}
		K(\theta)\vartheta^\mathrm{(h)}(\theta)\sim\frac{\vartheta^\mathrm{(p)}(\theta)}{K(\theta)}\sim K(\theta).
	\end{equation}
	Then in the definition of the form factors we can distinguish two kinds of contributions: the first is those coming from the minimal thermal form factor, which we disregard since they do not contain any term proportional to the ordinary form factors; the second is given by the terms contained in (\ref{ShG2pTFF}), especially in post-quench steady state's expectation value of the operator. The latter can be expanded according to the LeClair--Mussardo formula
	\begin{equation}
		\label{TEValpha}
		\langle e^{\ii\alpha\phi}\rangle_\rho = \mathcal{G}(\alpha) + \int_{-\infty}^{\infty}\frac{\dd\xi}{2\pi}|K(\xi)|^2F_{2;\mathrm{conn}}^\alpha(\xi) + O(K^4).
	\end{equation}
	If we now insert this into \eqref{1p/hContr} and use \eqref{ShG2pTFF}, we obtain terms containing 
	\begin{eqnarray}
		\label{FF2ndO}
		\int_{-\infty}^{\infty}\frac{\dd\xi}{2\pi}F_2^{\alpha}(-\theta,\theta)F_{2;\mathrm{conn}}^\alpha(\xi) = 
		\int_{-\infty}^{\infty}\frac{\dd\xi}{2\pi}F_{4;\mathrm{reg}}^\alpha(\xi+\ii\pi,-\theta,\theta,\xi).
	\end{eqnarray}
	The proof of this property is contained in the following subsection. It should be stressed that in this regime all the thermal form factors became vacuum ones. 
	
	\subsubsection{Proof of form factor decomposition (\ref{FF2ndO})}
	\label{Proof of Form Factor decomposition}
	The proof strongly relies on the clustering property (\ref{Cluster:Ax}), which is valid both for thermal and vacuum form factors. First one can apply the clustering property as
	\begin{eqnarray}
		F_2^{\alpha}(-\theta,\theta)F_\mathrm{2;conn}^\alpha(\xi) = \lim_{\Lambda\to\infty}F_\mathrm{4;reg}^\alpha(\xi+\Lambda+\ii\pi,-\theta,\theta,\xi+\Lambda) ,
	\end{eqnarray}
	then shift the integration variable $\xi$ by the constant $\Lambda$, getting back the following regular part of the form factor:
	\begin{equation}
		F_{4;\mathrm{reg}}^\alpha(\xi+\ii\pi,-\theta,\theta,\xi).
	\end{equation}
	A couple of subtleties must be discussed: first of all, the fact that there are other smooth functions in our integral does not change the main result (we push and pull back to the original variables in the end). Second, the clustering property links the same operators (ie, $\mathcal{O}=\mathcal{O}_1=\mathcal{O}_2$) and has an unit constant (ie, $\mathcal{N}^{\mathcal{O}}=1$) only in the case of vertex operators $\mathcal{O}=e^{\ii\alpha\phi}$ in the sinh-Gordon theory. So this is a very specific result, but, as we stated before, it can be readily generalised to other local fields in the sinh-Gordon model, even though they have different clustering properties. 
	
	\section{Semi-local sinh-Gordon operator's time evolution}
	\label{Semi-local Sinh-Gordon operator's time evolution}
	In this section we compute time evolution of $\tau$'s one-point function as usual by the means of the linked cluster expansion and the quench action method. 	Comparing the two we get the final expression (\ref{Tau}).
	
	Let us start with the linked cluster expansion; as previously outlined in Section \ref{Linked Cluster Expansion}, we do not have to consider contributions coming from stationary parts. We thus start expanding the initial state (\ref{InitialState}) and regularising it in a finite volume $L$. 	The first terms read
	\begin{eqnarray}
		C^\tau_{0,0}(t,L) &= \langle 0|\tau|0\rangle_L= \bar{\tau}, \\
		C^\tau_{0,2}(t,L) &= \sum_{I\in\mathrm{R}}  K(I)\frac{F_2^\tau(-\theta,\theta)}{\rho_2(-\theta,\theta;L)} e^{-2\ii mt \cosh\theta},\\
		C^\tau_{2,2}(t,L) &= \sum_{I,J\in \mathrm{R}} K^*(I)K(J)\langle I,-I|\tau|-J,J\rangle_L e^{2\ii mt(\cosh\xi - \cosh\theta)}\nonumber\\*
		&\qquad +C^\tau_{0,0}(t,L)\sum_{I\in \mathrm{R}}|K(I)|^2.		\label{RegCstau}
	\end{eqnarray} 
	We can express the finite-volume form factors in (\ref{RegCstau}) as
	\begin{eqnarray}
		\label{C22tau}
		C^\tau_{2,2}(t,L)&= \bar{\tau}\sum_{I\in \mathrm{R}}|K(I)|^2 \\
		&\hspace{-10mm}+ \sum_{I,J\in \mathrm{R}}  K^*(I)K(J)\frac{F_4^\tau(\xi+\ii\pi,-\xi+\ii\pi,-\theta,\theta)}{\rho_2(-\xi,\xi;L)\rho_2(-\theta,\theta;L)} e^{2\ii mt(\cosh\xi - \cosh\theta)}, \nonumber
	\end{eqnarray}
	and then split regular and singular parts as usual in a large but finite volume,
	\begin{eqnarray}
		\label{C22tau'}
		C^{\tau}_{2,2} (t,L\to\infty)&=& mL\bar{\tau}\int_{0}^{\infty}\frac{\dd\xi}{2\pi}|K(\xi)|^2 \cosh\xi - \Gamma t \bar{\tau} \nonumber\\
		&&+\int_{0}^{\infty}\frac{\dd\xi}{2\pi}\frac{d\theta}{2\pi}K^*(\xi)K(\theta)\nonumber\\*
		&&\qquad\times F_{4,\mathrm{reg}}^{\tau}(\xi+\ii\pi,-\xi+\ii\pi,-\theta,\theta)e^{2\ii mt(\cosh\xi - \cosh\theta)} \nonumber\\
		&\equiv& -Z_2\bar{\tau} - \Gamma t D^{\tau}_{0,0} + D^{\tau}_{2,2;\mathrm{conn}}(t). 
	\end{eqnarray}
	The linear term in $L$ will be cancelled since it can be rewritten as $\bar{\tau}Z_2$. Time evolution of the one-point function thus becomes
	\begin{eqnarray}
		\label{LCtau}
		\langle\tau(t)\rangle_\psi &=& (1-\Gamma t)\bar{\tau} +
		\int_{0}^{\infty}\frac{\dd\theta}{2\pi}K(\theta)F_2^\tau(-\theta,\theta)e^{-2\ii mt \cosh\theta}\\
		&&+\int_{0}^{\infty}\frac{\dd\xi}{2\pi}\frac{\dd\theta}{2\pi}K^*(\xi)K(\theta)\nonumber\\*
		&&\qquad\times F_{4;\mathrm{reg}}^{\tau}(\xi+\ii\pi,-\xi+\ii\pi,-\theta,\theta)e^{2\ii mt(\cosh\xi - \cosh\theta)} + \dots, \nonumber
	\end{eqnarray}
	in the linked cluster expansion up to the second order in $K(\theta)$.
	
	For the quench action method we simply note how the general treatment, discussed in \ref{Exponential decay from annihilation poles}, simplifies for the operator $\tau$. Specifically, the most divergent piece of the form factors can be expanded as
	\begin{eqnarray}
		\label{HOFFtaudiag}
		\langle\xi_1,-\xi_1,\dots,\xi_N,-\xi_N|&\tau|-\rho_N,\rho_N\dots,-\rho_1,\rho_1\rangle \nonumber\\*
		&\sim \bar{\tau}\prod_{b=1}^{N} \left[\frac{4\ii}{\left(\xi_{\sigma_a(b)} - \rho_b\right)^2} - \delta^2(\xi_{\sigma_a(b)} - \rho_b)\right]
	\end{eqnarray}
    for diagonal contributions, and
    \begin{eqnarray}
    	\label{HOFFtaunondiag}
    	&\langle\xi_1,-\xi_1,\dots,\xi_M,-\xi_M|\tau|-\rho_N,\rho_N\dots,-\rho_1,\rho_1\rangle \nonumber\\
    	&\qquad\sim \sum_{a_1,\dots,a_n=1}^{|N-M|}F_{2|N-M|}^{\tau}(\xi_{a_1},-\xi_{a_1},\dots,\xi_{a_p},-\xi_{a_p})\nonumber\\*
    	&\qquad\qquad\times\prod_{b=1}^{\min(N,M)} \left[\frac{4\ii}{\left(\xi_{\sigma_a(b)} - \rho_b\right)^2} - \delta^2(\xi_{\sigma_a(b)} - \rho_b)\right]
    \end{eqnarray}
    for off-diagonal terms with $p=|N-M|$. The diagonal terms then contribute at lowest order in $K$ as (\ref{HOFFtaudiag}), leading to the final result 
    \begin{equation}
    	f(t) = \bar{\tau} + \dots,
    \end{equation}
	where the dots are higher orders in powers of $K$, like the off-diagonal parts as well. 
		
	\section{Time evolution of local operators in the sine-Gordon model}
	\label{Local Sine-Gordon operator's time evolution}
	In this section we compute time evolution of the local vertex operator $e^{\ii\beta\phi}$ over the initial state (\ref{InitialState}) in the sine-Gordon model in its repulsive regime. This case differs from the previous ones as it contains different particle species, solitons ($+$) and antisolitons ($-$) respectively. We perform our computations using both linked cluster expansion and quench action method to check the formal resummation. The form factors of vertex operators in the sine-Gordon theory were derived in~\cite{Lukyanov:1997}. In this setting, first orders are retrieved straightforwardly as 
	\begin{eqnarray}
		\label{RegCsbeta}
		&C^\beta_{0,0}(t,L\to\infty) = \lim_{L\to\infty}\langle 0|e^{\ii\beta\phi}|0\rangle_L= \mathcal{G}_\beta, \\
		&C^\beta_{0,2}(t,L\to\infty) = \int_{0}^{\infty}\frac{\dd\theta}{2\pi}K^{ab}(\theta)F_{2,ab}^\beta(-\theta,\theta) e^{-2\ii mt \cosh\theta},
	\end{eqnarray}
	where we sum over recurring indices. The second order in $K(\theta)$ is more involved; we start from the formal expansion
	\begin{eqnarray}
		C^\beta_{2,2}(t,L) = \sum_{I,J\in R}&\mathcal{N}_{2,ab}(I,-I)\mathcal{N}_{2,cd}(-J,J) K^{ab,*}(I)K^{cd}(J)\nonumber\\*
		&\times\mbox{ }_{ab}\langle I,-I|e^{\ii\beta\phi}|-J,J\rangle_{cd} e^{2\ii mt(\cosh\xi - \cosh\theta)}
	\end{eqnarray}
	and then we expand the matrix element according to Smirnov's decomposition
	\begin{eqnarray}
		\label{SDbeta22}
		&\mbox{ }_{ab}\langle I,-I|e^{\ii\beta\phi}|-J,J\rangle_{cd} = \mathcal{G}_\beta\delta_{ac}\delta_{bd}\delta_{IJ}^2 \nonumber\\
		&\quad+ \delta_{ac}\delta_{IJ}F^\beta_{2,\bar{b}d}(\xi+\ii\pi,\theta) + \delta_{eg}[S_{ac}^{ef}(2\xi)]^* S_{bd}^{gh}(2\theta)\delta_{IJ}F^\beta_{2,\bar{f}h}(-\xi+\ii\pi,-\theta) \nonumber\\
		&\quad+ F_{4,\bar{a}\bar{b}cd}^\beta(\xi+\ii\pi,-\xi+\ii\pi,-\theta,\theta).
	\end{eqnarray}
	We can identify three kinds of contributions: disconnected (first line), semi-connected (second line), and fully connected (third line). We are going to discuss them separately; however, we stress as before that stationary contributions are only arising from semi-disconnected parts; they can be rewritten as 
	\begin{equation}
		\label{SCbeta22}
		2F^\beta\left[\sum_{I}K^{ab,*}(I)K_{ab}(I)\right],
	\end{equation}
	where we have used the following properties of scattering matrix
	\begin{equation}
	S_{ab}^{cd}(\xi)]^* = S_{ab}^{cd}(\xi),\quad S_{ab}^{ef}(-\xi)S_{ef}^{cd}(\xi) = \delta_{ab}^{cd}, \quad S_{ab}^{cd}(\xi) = S_{cd}^{ab}(\xi), 
	\end{equation}
	for $\xi\in\mathbb{R}$, and set~\cite{Lukyanov:1997} $F_{2,\bar{a}b}^\beta(\ii\pi,0) = F^\beta\delta_{ab}$. The disconnected terms provide the usual contribution $\mathcal{G}_\beta Z_2$ and the connected one does not contain singularities; hence we can write
	\begin{eqnarray}
		\label{C22betaPoles}
		C^\beta_{2,2}(t,L) &= \mathcal{G}_\beta Z_2(L) + F^\beta\int_{-\infty}^{\infty}\frac{\dd\theta}{2\pi} K^{ab,*}(\theta)K_{ab}(\theta) \\
		 &\quad+\int_{-\infty}^{\infty}\frac{\dd\xi}{2\pi}\frac{\dd\theta}{2\pi}K^{ab,*}(\xi)K^{cd}(\theta)\nonumber\\*
		 &\qquad\times F^\beta_{4,\bar{a}\bar{b}cd}(\xi+\ii\pi+\ii\epsilon,-\xi+\ii\pi+\ii\epsilon,-\theta,\theta)e^{2\ii mt(\cosh\xi - \cosh\theta)} \nonumber. 
	\end{eqnarray}
	This result shows that the zero-order term does not exponentiate for this local operator, since there is no term growing linear in $t$. Furthermore, we can also compute the time dependent non-resumming terms as in Section \ref{Linked Cluster Expansion}, which in the infinite-volume limit read 
	\begin{eqnarray}
		\label{SCbeta}
		D^\beta_{2,2;\mathrm{stat}} &= \int_{0}^{\infty}\frac{\dd\xi}{2\pi}K^{ab,*}(\xi)K_{ab}(\xi)F_{1,a;\mathrm{conn}}^\beta(\xi),\\
		D^\beta_{4,4;\mathrm{stat}} & =\frac{1}{2}\int_{0}^{\infty}\frac{\dd\xi_1}{2\pi}\frac{\dd\xi_2}{2\pi}K^{ab,*}(\xi_1)K_{ab}(\xi_1)K^{ab,*}(\xi_2)K_{ab}(\xi_2)\nonumber\\*
		&\qquad\times F_{2,ab;\mathrm{conn}}^\beta(\xi_1,\xi_2).
	\end{eqnarray}
	On the other hand, pure power-law contributions are
	\begin{eqnarray}
		\label{PLCbeta}
		D^\beta_{2,2;\mathrm{reg}}(t) &=\int_{0}^{\infty}\frac{\dd\xi}{2\pi}\frac{\dd\theta}{2\pi}K^{ab,*}(\xi)K_{ab}(\xi)K^{cd}(\theta)\nonumber\\*
		&\qquad\times F^\beta_{4,\bar{a}acd;\mathrm{reg}}(\xi+\ii\pi,-\theta,\theta,\xi)e^{2\ii mt\cosh\theta},\\
		D^\beta_{4,4;\mathrm{reg}}(t) &=\frac{1}{2}\int_{0}^{\infty}\frac{\dd\xi_1}{2\pi}\frac{\dd\xi_2}{2\pi}\frac{\dd\theta_1}{2\pi}K^{ab,*}(\xi_1)K^{cd}(\theta_1)K^{ef,*}(\xi_2)K_{ef}(\xi_2) \nonumber\\*
		&\qquad\times F_{6,\bar{a}\bar{b}\bar{e}cde;\mathrm{reg}}^\beta(\xi_1+\ii\pi,-\xi_1+\ii\pi,\xi_2+\ii\pi,-\theta_1,\theta_1,\xi_2)\nonumber\\*
		&\qquad\times e^{2\ii mt(\cosh\xi_1-\cosh\theta_1)},
	\end{eqnarray}
	where the first line provides a good approximation in the small-quench limit.
	
	Following~\cite{BSE14} and \ref{Exponential decay from annihilation poles} we can also get a resummed expression using the quench action method. The most divergent part of the form factors (\ref{HOFF}) in the diagonal terms reads
	\begin{eqnarray}
		\label{HOFFSGdiag}
		&\langle\xi_1,-\xi_1,\dots,\xi_N,-\xi_N|e^{\ii\beta\phi}|-\rho_N,\rho_N,\dots, -\rho_1,\rho_1\rangle\\
		&\quad\sim F_{4,\bar{a}\bar{b}cd;\mathrm{reg}}^{\beta}(\xi+\ii\pi,-\xi+\ii\pi,-\rho,\rho)\prod_{b=1}^{N-1} \left[\frac{4\ii}{\left(\xi_{\sigma_a(b)} - \rho_b\right)^2} - \delta^2(\xi_{\sigma_a(b)} - \rho_b)\right],\nonumber
	\end{eqnarray}
	while for off diagonal terms we get
	\begin{eqnarray}
		\label{HOFFSGnondiag}
		&\langle\xi_1,-\xi_1,\dots,\xi_{N+1},-\xi_{N+1}|e^{\ii\beta\phi}|-\rho_N,\rho_N,\dots, -\rho_1,\rho_1\rangle\\
		&\quad\sim F_{2;\bar{a}\bar{b},\mathrm{reg}}^{\beta}(\xi+\ii\pi,-\xi+\ii\pi)\prod_{b=1}^{N} \left[\frac{4\ii}{\left(\xi_{\sigma_a(b)} - \rho_b\right)^2} - \delta^2(\xi_{\sigma_a(b)} - \rho_b)\right].\nonumber
	\end{eqnarray}
	Obviously this differs from the semi-local case for what regards the prefactor of the exponential damping; in our formula (\ref{Main}), we can thus conclude 
	\begin{eqnarray}
		\label{fbeta}
		f(t) &= &\int_{0}^{\infty}K^{ab}(\theta)F_{2,ab}^\beta(-\theta,\theta)e^{-2\ii mt\cosh\theta} \\ 
		&&\quad + \int_{-\infty}^{\infty}\frac{\dd\xi}{2\pi}\frac{\dd\theta}{2\pi}K^{ab,*}(\xi)K^{cd}(\theta)\nonumber\\*
		&&\qquad\times F^\beta_{4,\bar{a}\bar{b}cd;\mathrm{reg}}(\xi+\ii\pi,-\xi+\ii\pi,-\theta,\theta)e^{2\ii mt(\cosh\xi - \cosh\theta)}. \nonumber
	\end{eqnarray}
	The power-law term $g(t)$ can in principle be derived using thermal form factors; for completeness we here just state the linked cluster expansion result
	\begin{eqnarray}
		\label{gbeta}
		g(t) &= \int_{0}^{\infty}\frac{\dd\xi}{2\pi}\frac{\dd\theta}{2\pi}K^{ab,*}(\xi)K_{ab}(\xi)K^{cd}(\theta)\nonumber\\*
		&\qquad\times F^\beta_{4,\bar{a}acd;\mathrm{reg}}(\xi+\ii\pi,-\theta,\theta,\xi)e^{-2\ii mt\cosh\theta}
	\end{eqnarray}
	in small quench regime. This term, when expanded and treated in a saddle point approximation, decays at late times as $g(t)\sim t^{-3/2}$.

	\section*{References}
	

\begin{thebibliography}{10}
		\bibitem{Deutsch-91}
		J. M. Deutsch, \emph{Quantum statistical mechanics in a closed system}, Phys. Rev. A \textbf{43}, 2046 (1991).
		
		\bibitem{Srednicki-99}
		M. Srednicki, \emph{The approach to thermal equilibrium in quantized chaotic systems}, J. Phys. A: Math. Gen. \textbf{32}, 1163 (1999).
		
		\bibitem{TurnerNature-18}
		C. Turner, A. Michailidis, D. Abanin, M. Serbyn and Z. Papic, \emph{Weak ergodicity breaking from quantum many-body scars}, Nat. Phys. \textbf{14}, 745 (2018).
		
		\bibitem{TurnerPRB-18}
		C. Turner, A. Michailidis, D. Abanin, M. Serbyn and Z. Papic, \emph{Quantum scarred eigenstates in a Rydberg atom chain: Entanglement, breakdown of thermalization, and stability to perturbations}, Phys. Rev. B 98 \textbf{15}, 155134 (2018).
		
		\bibitem{Sala-20}
		P. Sala, T. Rakovszky, R. Verresen, M. Knap and F. Pollmann, \emph{Ergodicity breaking arising from Hilbert space fragmentation in dipole-conserving Hamiltonians}, Phys. Rev. X 10 \textbf{1}, 011047 (2020).
		
		\bibitem{Moudgalya-22}
		S. Moudgalya, B. A. Bernevig, N. Regnault, \emph{Quantum many-body scars and Hilbert space fragmentation: A review of exact results}, Rep. Prog. Phys. \textbf{85}, 086501 (2022).
		
		\bibitem{Rigol-07}
		M. Rigol, V. Dunjko, V.Yurovsky and M. Olanshii, \emph{Relaxation in a completely integrable many-body quantum system: An ab initio study of the dynamics of the highly excited states of 1D lattice hard-core bosons}, Phys. Rev. Lett. \textbf{98}, 050405 (2007).
		
		\bibitem{Wouters-14}
		B. Wouters, J. De Nardis, M. Brockmann, D. Fioretto, M. Rigol and J.-S. Caux, \emph{Quenching the anisotropic Heisenberg chain: Exact solution and generalized Gibbs ensemble predictions}, Phys. Rev. Lett. \textbf{113}, 117202 (2014).
		
		\bibitem{Pozsgay-14}
		B. Pozsgay, M. Mestyán, M. A. Werner, M. Kormos, G. Zaránd and G. Takács, \emph{Correlations after quantum quenches in the XXZ spin chain: Failure of the generalized Gibbs ensemble}, Phys. Rev. Lett. \textbf{113}, 117203 (2014).
		
		\bibitem{Mussardo10}
		G.~Mussardo, \emph{{S}tatistical {F}ield {T}heory} (Oxford University Press,
		Oxford, 2010).
		
		\bibitem{Smirnov92book}
		F.~A. Smirnov, \emph{{F}orm {F}actors in {C}ompletely {I}ntegrable {M}odels of
			{Q}uantum {F}ield {T}heory} (World Scientific, Singapore, 1992).
		
		\bibitem{CalabreseCardy06}
		P.~Calabrese and J.~Cardy, \emph{{T}ime dependence of correlation functions
			following a quantum quench}, Phys. Rev. Lett. \textbf{96},  136801 (2006).
		
		\bibitem{Kinoshita-06}
		T. Kinoshita, T. Wenger and D. S. Weiss, \emph{A quantum Newton's cradle}, Nature \textbf{440}, 900 (2006).
		
		\bibitem{Gring-12}
		M. Gring, M. Kuhnert, T. Langen, T. Kitagawa, B. Rauer, M. Schreitl, I. Mazets, D. A. Smith, E. Demler and J. Schmiedmayer, \emph{Relaxation and prethermalization in an isolated quantum system}, Science \textbf{337}, 1318 (2012).

		\bibitem{Langen-15}
		T. Langen, S. Erne, R. Geiger, B. Rauer, T. Schweigler, M. Kuhnert, W. Rohringer, I. E. Mazets, T. Gasenzer and J. Schmiedmayer, \emph{Experimental observation of a generalized Gibbs ensemble}, Science \textbf{348}, 207 (2015).
		
		\bibitem{Geiger-14}
		R. Geiger, T. Langen, I.E. Mazets and J. Schmiedmayer, \emph{Local relaxation and light-cone-like propagation of correlations in a trapped one-dimensional Bose gas}, New J. Phys. \textbf{5}, 053034 (2014).
		
		\bibitem{AduSmith-13}
		D. Adu Smith, M. Gring, T. Langen, M. Kuhnert, B. Rauer, R. Geiger, T. Kitagawa, I. Mazets, E. Demler and J. Schmiedmayer. \emph{Prethermalization revealed by the relaxation dynamics of full distribution functions}, New J. Phys. \textbf{15}, 075011 (2013).
		
		\bibitem{Kitagawa-11}
		T. Kitagawa, A. Imambekov, J. Schmiedmayer and E. Demler, \emph{The dynamics and prethermalization of one-dimensional quantum systems probed through the full distributions of quantum noise}, New J. Phys. \textbf{13}, 073018 (2011).
		
		\bibitem{Niklas-15}
		E. Nicklas, M. Karl, M. Höfer, A. Johnson, W. Muessel, H. Strobel, J. Tomkovič, T. Gasenzer and M. K. Oberthaler, \emph{Observation of scaling in the dynamics of a strongly quenched quantum gas}, Phys. Rev. Lett. \textbf{24}, 245301 (2015).
		
		\bibitem{Meinert-15}
		F. Meinert, M. Panfil, M. J. Mark, K. Lauber, J.-S. Caux and H.-C. Nägerl, \emph{Probing the excitations of a Lieb-Liniger gas from weak to strong coupling}, Phys. Rev. Lett. \textbf{8}, 085301 (2015).
		
		\bibitem{Meinert-14}
		F. Meinert, M. J. Mark, E. Kirilov, K. Lauber, P. Weinmann, M. Gröbner, A. J. Daley and H.-C. Nägerl, \emph{Observation of many-body dynamics in long-range tunneling after a quantum quench}, Science \textbf{344}, 6189 (2014).
		
		\bibitem{Malvania-21}
		N. Malvania, Y. Zhang, Y. Le, J. Dubail, M. Rigol and D.S. Weiss, \emph{Generalized hydrodynamics in strongly interacting 1D Bose gases}, Science \textbf{6559}, 1129 (2021).	
		
		\bibitem{ZamolodchikovZamolodchikov79}
		A.~B. Zamolodchikov and Al.~B. Zamolodchikov, \emph{Factorized S-matrices in two dimensions as the exact solutions of certain relativistic quantum field theory models}, Ann. Phys. {\bf 120}, 253 (1979).
		
		\bibitem{Faddeev80}
		L.~D. Faddeev, \emph{Quantum completely integrable models in field theory}, Sov. Sci. Rev. Math. Phys.~C {\bf 1}, 107 (1980).
		
		\bibitem{Parke80}
		S.~Parke, \emph{Absence of particle production and factorization of the S-matrix in 1+1 dimensional models}, 
		Nucl. Phys. B \textbf{174}, 166 (1980).
		
		\bibitem{Ginsparg:notes}
		P.~Ginsparg,
		\emph{{A}pplied {C}onformal {F}ield {T}heory}, Fields, Strings and Critical Phenomena (Les Houches, Session XLIX, 1988), ed. by E. Brézin and J. Zinn Justin, (1989).
		
		\bibitem{Cardy:08}
		J. L.~Cardy, O. A.Castro-Alvaredo, and B.~Doyon,
		\emph{{F}orm factors of branch-point twist fields in quantum integrable models and entanglement entropy}, J. Stat. Phys. \textbf{130}, 129 (2008).
		
		 \bibitem{YurovZamolodchikov91}
		V. P. Yurov and A. B. Zamolodchikov, \emph{Correlation functions of integrable 2-D models of relativistic field theory. Ising model}, Int. J. Mod. Phys. A \textbf{6}, 3419 (1991).
		
		\bibitem{Kadanoff:71}
		L. P.~Kadanoff and H.~Ceva,  
		\emph{Determination of an operator algebra for the two-dimensional Ising model}, Phys. Rev. B \textbf{3}, 3918 (1971).
		
		\bibitem{Polkovnikov-11}
		A.~Polkovnikov, K.~Sengupta, A.~Silva and M.~Vengalattore,
		\emph{{N}onequilibrium dynamics of closed interacting quantum systems}, Rev.
		Mod. Phys. \textbf{83},  863 (2011).
		
		\bibitem{FiorettoMussardo10}
		D.~Fioretto and G.~Mussardo, \emph{{Q}uantum quenches in integrable field
			theories}, New J. Phys. \textbf{12},  055015 (2010).
		
		\bibitem{Sotiriadis-12}
		S.~Sotiriadis, D.~Fioretto and G.~Mussardo, \emph{{Z}amolodchikov-{F}addeev
			algebra and quantum quenches in integrable field theories}, J. Stat. Mech.
		(2012) P02017.
		
		\bibitem{Cazalilla06}
		M. A.~Cazalilla, \emph{{E}ffect of Suddenly Turning on Interactions in the {L}uttinger Model}, Phys. Rev. Lett. \textbf{97}, 156403 (2006).
		
		\bibitem{Rossini-10}
		D.~Rossini, S.~Suzuki, G.~Mussardo, G.~E. Santoro and A.~Silva, \emph{{L}ong
			time dynamics following a quench in an integrable quantum spin chain: {L}ocal
			versus nonlocal operators and effective thermal behavior}, Phys. Rev. B
		\textbf{82},  144302 (2010).
		
		\bibitem{GhoshalZamolodchikov94}
		S.~Ghoshal and A.~B. Zamolodchikov, \emph{{B}oundary {S}-matrix and boundary
			state in two-dimensional integrable quantum field theory}, Int. J.~Mod.
		Phys.~A \textbf{9},  3841 (1994); \emph{ibid.} \textbf{9}, 4353(E) (1994).
		
		\bibitem{CS17}
		A.~Cort\'{e}s~Cubero and D.~Schuricht, \emph{Quantum quench in the attractive
			regime of the sine-{G}ordon model}, J. Stat. Mech. (2017) 103106.
		
		\bibitem{Horvath-18}
		D.~X. Horv\'{a}th, M.~Kormos and G.~Tak\'{a}cs, \emph{Overlap singularity and time evolution in integrable quantum field theory}, JHEP \textbf{08}, 170 (2018).
		
		\bibitem{Calabrese-12jsm1}
		P.~Calabrese, F.~H.~L. Essler and M.~Fagotti, \emph{Quantum quench in the transverse field Ising chain: I. Time evolution of order parameter correlators}, J. Stat. Mech. (2012) P07016.
		
		\bibitem{Calabrese-12jsm2}
		P.~Calabrese, F.~H.~L. Essler and M.~Fagotti, \emph{Quantum quench in the transverse field Ising chain: II. Stationary state properties}, J. Stat. Mech. (2012) P07022.
		
		\bibitem{Delfino14}
		G.~Delfino, \emph{{Q}uantum quenches with integrable pre-quench dynamics}, J.
		Phys. A: Math. Theor. \textbf{47},  402001 (2014).
		
		\bibitem{DelfinoViti17}
		G.~Delfino and J.~Viti, \emph{On the theory of quantum quenches in
			near-critical systems}, J. Phys. A: Math. Theor. \textbf{50},  084004 (2017).
		
		\bibitem{SE12}
		D.~Schuricht and F.~H.~L. Essler, \emph{{D}ynamics in the {I}sing field theory
			after a quantum quench}, J.~Stat. Mech. (2012) P04017.
		
		\bibitem{BSE14}
		B.~Bertini, D.~Schuricht and F.~H.~L. Essler, \emph{{Q}uantum quench in the
			sine-{G}ordon model}, J.~Stat. Mech. (2014) P10035.
		
		\bibitem{PozsgayTakacs08-1}
		B.~Pozsgay and G.~Tak\'{a}cs, \emph{Form factors in finite volume {I: Form}
			factor bootstrap and truncated conformal space}, Nucl. Phys. B \textbf{788},
		167 (2008).
		
		\bibitem{PozsgayTakacs08-2}
		B.~Pozsgay and G.~Tak\'{a}cs, \emph{Form factors in finite volume {II: Disconnected} terms and finite
			temperature correlators}, Nucl. Phys. B \textbf{788},  209 (2008).
		
		\bibitem{LeclairMussardo99}
		A.~LeClair and G.~Mussardo, \emph{{F}inite temperature correlation functions in
			integrable {QFT}}, Nucl. Phys.~B \textbf{552},  624 (1999).
		
		\bibitem{KormosPozsgay10}
		M.~Kormos and B.~Pozsgay, \emph{{O}ne-point functions in massive integrable QFT with boundaries}, JHEP \textbf{04}, 112 (2010).
		
		\bibitem{DiSalvo:23}
		E.~Di Salvo and D.~Schuricht,
		\emph{{Q}uantum quenches in the sinh-{G}ordon and {L}ieb–{L}iniger models}, J. Stat. Mech. (2023) 053107.
		
		\bibitem{CauxEssler13}
		J.-S. Caux and F.~H.~L. Essler, \emph{{T}ime evolution of local observables
			after quenching to an integrable model}, Phys. Rev. Lett. \textbf{110},
		257203 (2013).
		
		\bibitem{Caux16}
		J.-S. Caux, \emph{The quench action}, J. Stat. Mech. (2016) 064006.
		
		\bibitem{Doyon-23}
		B. Doyon, G. Perfetto, T. Sasamoto and T. Yoshimura, \emph{Ballistic macroscopic fluctuation theory}, SciPost Phys. \textbf{15}, 136 (2023).
		
		\bibitem{CuberoPanfil19}
		A.~Cort\'{e}s~Cubero and M.~Panfil, \emph{{T}hermodynamic bootstrap program for integrable QFT’s: form factors and correlation functions at finite energy density}, J. High Energ. Phys \textbf{2019}, 104 (2019).
		
		\bibitem{CuberoPanfil20}
		A.~Cort\'{e}s~Cubero and M.~Panfil, \emph{{G}eneralized hydrodynamics regime 
			from the thermodynamic bootstrap program}, SciPost Phys. \textbf{8}, 004 (2020).
		
		\bibitem{Granet-20}
		E.~Granet, M.~Fagotti and F.~H.~L.~Essler, \emph{{F}inite temperature and quench dynamics in the transverse field {I}sing model from form factor expansions}, SciPost Phys. \textbf{9}, 033 (2020).
		
		\bibitem{Its93}
		A. R. ~Its, A. G.~Izergin, V. E.~Korepin and  N. A~Slavnov, \emph{{T}emperature correlations of quantum spins}, Phys. Rev. Lett. \textbf{70}, 11 (1993).
		
		\bibitem{Konik-21}
		R. M.~Konik, M.~L\'{a}jer and G.~Mussardo, \emph{Approaching the self-dual point
			of the sinh-{G}ordon model}, J. High Energ. Phys. \textbf{2021},  14 (2021).
		
		\bibitem{LukyanovZamolodchikov97}
		S.~Lukyanov and A.~B. Zamolodchikov, \emph{{E}xact expectation values of local
			fields in the quantum sine-{G}ordon model}, Nucl. Phys.~B \textbf{493},  571
		(1997).
		
		\bibitem{Essler:Review}
		F. H. L.~Essler and R. M.~Konik, in \emph{{F}rom {F}ields to {S}trings: {C}ircumnavigating {T}heoretical {P}hysics}, ed.~M.~Shifman, A.~Vainshtein and J.~Wheater (World Scientific, Singapore, 2005).
		
		\bibitem{Giamarchi04}
		T.~Giamarchi,
		\emph{{Q}uantum {P}hysics in {O}ne {D}imension} (Oxford University Press, Oxford, 2004).
		
		\bibitem{Fendley:95}
		P.~Fendley, A. W. W.~Ludwig, and H.~Saleur,
		\emph{{E}xact nonequilibrium transport through point contacts in quantum wires and fractional quantum {H}all devices}, Phys. Rev. B \textbf{52}, 12 (1995).
		
		\bibitem{KorepinBogoliubovIzergin93}
		V.~E. Korepin, N.~M. Bogoliubov and A.~G. Izergin,
		\emph{{Q}uantum {I}nverse {S}cattering {M}ethod and {C}orrelation {F}unctions} (Cambridge University
		Press, Cambridge, 1997).
		
		\bibitem{Coleman:75}
		S.~Coleman,
		\emph{Quantum sine-Gordon equation as the massive Thirring model}, Phys. Rev. D \textbf{11}, 8 (1975).
		
		\bibitem{Barmettler-09}
		P.~Barmettler, M.~Punk, V.~Gritsev, E.~Demler, and E.~Altman,
		\emph{{R}elaxation of antiferromagnetic order in spin-1/2 chains following a quantum quench}, Phys. Rev. Lett. \textbf{102}, 130603 (2009)
		
		\bibitem{Barmettler-10}
		P.~Barmettler, M.~Punk, V.~Gritsev, E.~Demler, and E.~Altman, \emph{Quantum quenches in the anisotropic spin-1/2 Heisenberg chain: different approaches to many-body dynamics far from equilibrium}, New J. Phys. \textbf{12}, 055017 (2010).
		
		\bibitem{KormosZarand15}
		M.~Kormos and G.~Zar\'and,
		\emph{Quantum quenches in the sine-Gordon model: A semiclassical approach}, Phys. Rev. E. \textbf{93}, 6 (2015).
		
		\bibitem{MocaKormosZarand15}
		P.~Moca, M.~Kormos, and G.~Zar\'and,
		\emph{{H}ybrid semiclassical theory of quantum quenches in one-dimensional systems}, Phys. Rev. Lett. \textbf{119}, 100603 (2017).
		
		\bibitem{Bertini:16}
		B.~Bertini, M.~Collura, J.~De Nardis and M.~Fagotti,
		\emph{{T}ransport in out-of-equilibrium {XXZ} chains: {E}xact profiles of charges and currents}, Phys. Rev. Lett. \textbf{117}, 207201 (2016).
		
		\bibitem{CastroAlvaredo:16}
		O. A.~Castro-Alvaredo, B.~Doyon and T.~Yoshimura,
		\emph{{E}mergent hydrodynamics in integrable quantum systems out of equilibrium}, Phys. Rev. X \textbf{6}, 041065 (2016).
		
		\bibitem{Doyon:23}
		B.~Doyon, G.~Perfetto, T.~Sasamoto and T.~Yoshimura,
		\emph{{B}allistic macroscopic fluctuation theory}, SciPost Phys. \textbf{15}, 136 (2023).
		
		\bibitem{Delfino04}
		G.~Delfino, \emph{{I}ntegrable field theory and critical phenomena: the {I}sing model in a magnetic field}, 
		J.~Phys.~A: Math. Gen. \textbf{37}, R45 (2004).	
		
		\bibitem{Horvath-20}
		D.~X. Horv\'{a}th, and P.~Calabrese, \emph{{S}ymmetry resolved entanglement in integrable field theories via form factor bootstrap}, JHEP \textbf{11}, 131 (2020).
		
		\bibitem{KoubekMussardo93}
		A.~Koubek and G.~Mussardo, \emph{{O}n the operator content of the sinh-{G}ordon
			model}, Phys. Lett.~B \textbf{311},  193 (1993).
		
		\bibitem{Fring-93}
		A.~Fring, G.~Mussardo and P.~Simonetti, \emph{{F}orm factors for integrable
			{L}agrangian field theories, the sinh-{G}ordon model}, Nucl. Phys.~B
		\textbf{393},  413 (1993).
		
		\bibitem{CardyMussardo90}
		J.~L. Cardy and G. Mussardo, \emph{{F}orm factors of descendent operators in perturbed conformal field theories}, Nucl. Phys. B \textbf{340}, 387 (1990).
		
		\bibitem{Delfino:96}
		G.~Delfino, P.~Simonetti, and J. L.~Cardy,
		\emph{Asymptotic factorisation of form factors in two-dimensional quantum field theory}, Phys. Lett. B \textbf{387}, 2 (1996).
		
		\bibitem{GradshteynRyzhik80}
		I.~S. Gradshteyn, and I.~M. Ryzhik,
		\emph{{T}able of {I}ntegrals, {S}eries, and {P}roducts} (Academic Press, London, 1980).
		
		\bibitem{Essler24}
		F.~H.~L.~Essler, and A.~J.~J.~M.~de Klerk,
		\emph{{S}tatistics of matrix elements of local operators in integrable models}, Phys. Rev. X \textbf{14}, 031048 (2024).
		
		\bibitem{GranetEssler21}
		E. Granet and F. H. L. Essler, \emph{Systematic strong coupling expansion for out-of-equilibrium dynamics in the Lieb-Liniger model}, SciPost Phys. \textbf{11}, 068 (2021).
		
		\bibitem{PanfilKonik23}
		M.~Panfil and R.~M. Konik, \emph{{E}xtending the thermodynamic form factor bootstrap program: {M}ultiple particle-hole excitations, crossing symmetry, and reparameterization invariance}, J. High Energ. Phys., \textbf{72} (2023).
		
		\bibitem{DeNardisCaux14}
		J.~De~Nardis and J.~S. Caux, \emph{{A}nalytical expression for a post-quench
			time evolution of the one-body density matrix of one-dimensional hard-core
			bosons}, J. Stat. Mech. (2014) P12012.
		
		\bibitem{DeNardis-14}
		J.~{De Nardis}, B.~Wouters, M.~Brockmann and J.-S. Caux, \emph{{S}olution for
			an interaction quench in the {Lieb-Liniger Bose} gas}, Phys. Rev. A
		\textbf{89},  033601 (2014).
		
		\bibitem{DeNardis-15}
		J.~De~Nardis, L.~Piroli and J.-S. Caux, \emph{{R}elaxation dynamics of local
			observables in integrable systems}, J. Phys. A: Math. Theor. \textbf{48},
		43FT01 (2015).
		
		 \bibitem{Doyon07-2}
		B. Doyon, \emph{Finite-temperature form factors: a review}, SIGMA \textbf{3}, 011 (2007).
		
		\bibitem{LeClair:96}
		A.~LeClair, F.~Lesage, S.~Sachdev, and H.~Saleur,
		\emph{{F}inite temperature correlations in the one-dimensional quantum Ising model}, Nucl. Phys. B \textbf{482}, 3 (1996).
		
		\bibitem{Doyon:05}
		B.~Doyon,
		\emph{Finite-temperature form factors in the free Majorana theory}, J. Stat. Mech. (2005) P11006.
		
		\bibitem{Lukyanov:1997}
		S.~Lukyanov,
		\emph{{F}orm-factors of exponential fields in the sine-{G}ordon model}, Mod. Phys. Lett. A \textbf{12}, 2543 (1997).
			
	\end{thebibliography}

\end{document}